\documentclass[fleqn,usenatbib]{mnras}

\usepackage{newtxtext,newtxmath}
\usepackage[varg]{txfonts}
\usepackage[T1]{fontenc}
\usepackage{ae,aecompl}
\usepackage{natbib,twoopt}
\usepackage{graphicx}	% Including figure files
\usepackage{amsmath}	% Advanced maths commands
\usepackage{amssymb}	% Extra maths symbols
\usepackage{amsfonts}
\usepackage{wasysym}
\usepackage{pdflscape}
\usepackage{comment}
\usepackage{float}
\usepackage{csquotes}

\title[A census of young stars associated with HD 200775]{A census of young stellar population associated with the Herbig Be star HD 200775}

\author[Piyali Saha et al.]{Piyali Saha$^{1,2}$\thanks{E-mail: s.piyali16@gmail.com (PS)},
Maheswar, G$^{1}$,
Kamath, U. S.$^{1}$,
Lee, C. W$^{3}$,
Manoj, P.$^{4}$,
\newauthor{Blesson Mathew$^{5}$,
Ekta Sharma$^{1}$}
\\
$^{1}$Indian Institute of Astrophysics (IIA), Sarjapur Road, Koramangala, Bangalore 560034, India\\
$^{2}$Pt. Ravishankar Shukla University, Amanaka G.E. Road, Raipur, Chhatisgarh 492010, India\\
$^{3}$Korea Astronomy, and Space Science Institute (KASI), 776 Daedeokdae-ro, Yuseong-gu, Daejeon 305-348, Republic of Korea\\
$^{4}$Tata Institute of Fundamental Research, Homi Bhabha Road, Mumbai 400 005, India\\
$^{5}$Department of Physics and Electronics, CHRIST (Deemed to be University), Bangalore 560029, India\\
}
\date{Accepted XXX. Received YYY; in original form ZZZ}

\pubyear{2020}

\begin{document}
\label{firstpage}
\pagerange{\pageref{firstpage}--\pageref{lastpage}}
\maketitle

% Abstract of the paper
\begin{abstract}  
The region surrounding the well-known reflection nebula, NGC 7023, illuminated by a Herbig Be star, HD 200775, located in the dark cloud L1174 is studied in this work. Based on the distances and proper motion values from \textit{Gaia} DR2 of 20 previously known young stellar object candidates, we obtained a distance of $335\pm11$ pc to the cloud complex L1172/1174. Using polarization measurements of the stars projected on the cloud complex, we show additional evidence for the cloud to be at $\sim335$ pc distance. Using this distance and proper motion values of the YSO candidates, we searched for additional comoving sources in the vicinity of HD 200775 and found 20 new sources which show low infrared excess emission and are of age $\sim1$ Myr. Among these, 10 YSO candidates and 4 newly identified comoving sources are found to show X-ray emission. Three of the four new sources for which we have obtained optical spectra show H$\alpha$ in emission. About 80\% of the total sources are found within $\sim1$ pc distance from HD 200775. Spatial correlation of some of the YSO candidates with the \textit{Herschel} dust column density peaks suggests that star formation is still active in the region and may have been triggered by HD 200775.  
\end{abstract}


%%%%%%%%%%%%%%%%%%%%%%%%%%%%%%%%%%%%%%%%%%%%%%%%%%

%%%%%%%%%%%%%%%%% BODY OF PAPER %%%%%%%%%%%%%%%%%%

\section{Introduction}\label{sec:intro}

A complete census of young stars and sub-stellar objects in the nearby star forming regions is essential to improve our understanding of the global properties of younger population like disc fraction, initial mass function, star formation efficiency and the star formation history. One of the key requirements to carry out such scientific investigations is the distance of the region from the Sun. It is an essential parameter required to determine many of the physically relevant properties such as the mass and the physical size of the molecular cloud. It is also required to determine the luminosity of young stellar objects (YSOs) that are embedded in the molecular cloud and the size of outflows, if present \citep{1990ApJ...365L..73Y,1991ApJS...75..877C,2008A&A...487..993K}. Unfortunately, the distance estimation of many of the clouds is often plagued by very large uncertainties and it is seen in the literature that measurements often differ by large factors, especially for those clouds that are isolated, sometimes by as much as a factor of two \citep{1995A&AS..113..325H}. 

LDN 1172/1174 \citep[hereafter L1172/1174; ][]{1962ApJS....7....1L} is an isolated star forming region situated at a relatively high Galactic latitude ($l\sim104.1\degr$, $b\sim+14.2\degr$). L1172/1174 is associated with a bright reflection nebula NGC 7023 in the Cepheus constellation. The nebula is illuminated by a Herbig Be (B2/3Ve) star HD 200775 \citep{1994A&AS..104..315T,2006ApJ...653..657M} which is identified as a double-line spectroscopic binary system having primary and secondary masses of 10.7$\pm$2.5 M$_{\odot}$ and 9.3$\pm$2.1 M$_{\odot}$ respectively \citep{2008MNRAS.385..391A}. Being associated with a single filament, the whole cloud system resembles a \textquote{head-tail} structure oriented at an angle of $\sim50$\degr~to the Galactic plane \citep{1978ApJ...220..510E}. An outflow from HD 200775, currently inactive though, is responsible for generating an asymmetric east-west biconical cavity that is filled with hot atomic gas \citep{1998A&A...339..575F}.   \cite{2013AJ....145...35R} discovered four new Herbig-Haro (HH) objects in NGC 7023 by wide-field imaging. At least two distinct outflows were discovered in the northwestern \textquote{lobe} of NGC 7023. \cite{2009ApJS..185..198K} identified a total of 50 YSO candidates in the vicinity of HD 200775 based on the \textit{Spitzer} Infrared Array Camera (IRAC) and the Multiband Imaging Photometer for \textit{Spitzer} (MIPS) study of the Cepheus Flare. HD 200775 is located at the center of a sparse cluster of T Tauri stars \citep[TTS; ][]{1953AJ.....58...48W}. Observationally, TTS are identified based on their proximity to molecular clouds and presence of Balmer lines of hydrogen in emission \citep{1989A&ARv...1..291A}. The equivalent width of H${\alpha}$ emission line, EW(H${\alpha}$), is used to classify the TTS to classical (CTTS) and weak-line (WTTS) sources. About 14 of the TTS found near HD 200775 show H$\alpha$ in emission and variability.  At least 3 of these TTS are found to be in binary systems \citep{1975Ap&SS..33..487R}. Based on $^{13}$CO observations, \cite{2013MNRAS.429..954Y} suggested that strong winds from HD 200775 are blowing away ambient material causing further compression of the matter around it, especially to the northern region where most of the YSOs are found to be distributed. They also found a systematic decrease in the age of YSOs as a function of distance from HD 200775 implying that at least some of them are possibly formed as a result of the feedback from HD 200775. Presence of a Herbig Be star, HH objects and low mass YSOs makes L1172/1174 an excellent candidate to study star formation possibly due to stellar feedback.

The distance to L1172/1174 is not firmly constrained. Several values of distances have been quoted in the literature. The distance estimates made for the complex are compiled from literature and listed in Table \ref{tab:dist_table}. The estimated distances range from $\sim300$ pc to $\sim700$ pc. Based on the extinction and the absolute magnitude of HD 200775, \cite{1969MmSAI..40...75V} made initial estimates of the distance to the star and obtained a value of 400$\pm$100 pc. Using photoelectric photometry and low resolution spectroscopy of 75 stars projected against L1172/1174, ~\citet{1989SvA....33..487S} reported a distance of 300 $\pm$ 20 pc by producing a colour-excess vs distance plot. \citet{1992BaltA...1..149S} used Vilnius photometry which gives two-dimensional classification and interstellar reddening produced extinction vs. distance plot of 79 stars towards the cloud. They obtained a distance of 288$\pm$25 pc to L1172/1174. The parallax measurements of HD 200775 were used to calculate a distance of 430$^{160}_{-90}$ pc to it \citep{1998A&A...330..145V}. However, based on a recomputed Hipparcos parallax value, \citet{2007A&A...474..653V} estimated a distance of 520$^{+180}_{-110}$ pc to HD 200775. \cite{2010A&A...509A..44M}, by using 2MASS $JHK$ photometry obtained extinction and distances of sources projected against the cloud, estimated a distance of 408$\pm$76 pc. \cite{2013A&A...555A.113B} obtained a revised distance of 320$\pm$51 pc to HD 200775 by combining radial velocity measurements and astrometric data. Using Wolf diagrams, \cite{1998ApJS..115...59K} suggested that the Cepheus region above the Galactic latitude of $+10\degr$ consists of different cloud material partly projected against each  other. They found evidence of absorbing clouds at $\sim200$ and $\sim400$ pc and assigned a distance of 450$\pm45$ pc to L1172/1174. Evidence for the presence of two layers of interstellar gas was found in neutral hydrogen \citep{1967ApJS...15...97H} and in CO molecular line \citep{1989ApJ...347..231G} observations. By applying kinematical method to velocity profiles of the lines, \citet{1989ApJ...347..231G} estimated an approximate distances of 300 pc and 800 - 900 pc to the layers. \cite{2009BaltA..18...33Z} investigated $1.5\degr\times1.5\degr$ area centering at $l$=$104.1\degr,b$=$+14.2\degr$. This area was divided into five smaller subareas and dependence of extinctions on distances were studied by using stars projected on these subareas. They suggested that the dust clouds in the vicinity of NGC 7023 are concentrated in at least two layers that are at $\sim280$ pc and $\sim715$ pc. However, in their analysis, the number of stars used to discern the hikes in extinction in different subareas is very small making the distance estimates quite uncertain.

\begin{table}
\caption{List of previous distance estimation of L1172/1174 compiled from the literature.}
	\label{tab:dist_table}
	\begin{tabular}{p{1cm}p{5.5cm}p{0.4cm}} % four columns, alignment for each
		\hline
        Distance   & Methods used to obtain the distance            &Ref.\\
        (pc)       &                                                &     \\
        \hline
        282$\pm$42$^{\dagger}$ & Extinction vs distance                         &1\\
        288$\pm$25 & Extinction vs distance                         &2\\
        300$\pm$20 & Color-excess vs distance                       &3\\
        320$\pm$51 & Radial velocity and astrometry of HD 200775    &4\\
        358$\pm$31 & Distances of YSOs  in Cepheus flare region     &5\\
        408$\pm$76 & Extinction vs distance ($JHK$ photometry)      &6\\
        430$^{160}_{-90}$ & $Hipparcos$ distance of HD 200775       &7 \\
        440$\pm$100 & Extinction and absolute magnitude of HD 200775    &8\\
        450$\pm$45  & Wolf diagrams                                 &9\\
        510$\pm$130 & Star counts method                            &10\\
        520$^{180}_{-110}$ & Recomputed $Hipparcos$ distance of HD 200775   &11\\
        715$\pm$110$^{\dagger}$&Extinction vs distance              &1\\
        \hline
	\end{tabular}\\
    
    1. \cite{2009BaltA..18...33Z}, 2. \cite{1992BaltA...1..149S}, 3. \citet{1989SvA....33..487S}, 4. \cite{2013A&A...555A.113B}, 5. \cite{2018ApJ...867..151D}, 6 \cite{2010A&A...509A..44M}, 7. \citet{1998A&A...330..145V}, 8. \cite{1969MmSAI..40...75V}, 9. \cite{1998ApJS..115...59K}, 10. \cite{1981AJ.....86.1923A}, 11. \cite{2007A&A...474..653V}\\
    $^{\dagger}$ \cite{2009BaltA..18...33Z} identified two layers of material towards the direction of L1172/1174 based on the jump in extinction seen in the extinction vs distance plot.
\end{table}
 
On the basis of the spectral slope of their spectral energy distribution (SED) in the near- to submillimeter wavelengths, the YSOs are also classified on an empirical sequence into Class 0 - III and flat spectrum sources  \citep{1987IAUS..115....1L, 1993ApJ...406..122A, 1994ApJ...434..614G, 2007ApJS..169..328R, 2009ApJS..181..321E, 2010ApJS..188...75M}. The Class II and Class III objects generally correspond to CTTS and WTTS, respectively. Based on the H$\alpha$ emission, near-IR and mid-IR observations, a total of $\sim60$ YSO candidates have been identified so far in NGC 7023 \citep{2009ApJS..185..451K, 2009ApJS..185..198K, 2013MNRAS.429..954Y}. Because WTTSs show low EW(H${\alpha}$) and less IR excess emission, it is difficult to distinguish them from the unrelated field stars and they may fail detection in H${\alpha}$ and IR surveys. Thus a complete census of Class III sources in a star forming region may not always be complete. Using proper motion and parallax measurements from \textit{Gaia} DR2, numerous kinematically coherent associations of stars have been recognized within a few hundred parsec of the Sun \citep[e.g., ][]{2018ApJ...867..151D, 2018AJ....156...76L, 2018ApJ...863...91F}. Detailed study of the nearest of these associations has revealed the presence of a large number of low mass stars and sub-stellar objects \citep{2018AJ....156..271L, 2018AJ....156...76L}. Thus, an alternative way to get census of disc-less young members, if present, in a star forming region is to look for additional sources that are kinematically associated with the known YSOs of the region. We made a search for any additional sources that may be comoving with the known YSO candidates and may have been missed detection in the earlier studies of L1172/1174 \citep{2009ApJS..185..451K, 2009ApJS..185..198K, 2013MNRAS.429..954Y}, using distance and proper motion values from the \textit{Gaia} DR2 database.

In this work we estimated the distance to LDN 1172/1174 using recently released \textit{Gaia} DR2 data of the YSO candidates identified in the direction of the cloud and of the sources for which we have made R-band polarimetric observations. Using the distances and the proper motion values of the YSO candidates as reference, we searched for additional co-moving sources in the cloud. We obtained a number of sources moving in a similar fashion as the YSO candidates and lying in a similar distance. We classified these comoving sources based on the additional information collated from the archives. The paper is organized in the following manner: details of our observations and the \textit{Spitzer} and the \textit{Gaia} DR2 data sets are described in section \ref{sec:obs}; results and discussion are presented in section \ref{sec:res_dis}. Finally, we conclude the paper with a summary of the results in section \ref{sec:con}.

%####################################################################################
\section{Observation and data reduction} \label{sec:obs}

\subsection{Polarization measurements}

The polarimetric observations of 42 fields covering the cloud L1172/1174 were carried out using ARIES Imaging POLarimeter (AIMPOL; \cite{2004BASI...32..159R}) mounted at the Cassegrain focus of 1.04m Sampurnanand Telescope, ARIES, Nainital, India. Observations were performed on 26 nights spanning over three years from 2015 to 2017 (see Table \ref{tab:log_table}). The frames were obtained using a 1024 $\times$ 1024 pixel$^{2}$ CCD chip (Tektronix TK1024), of which central 325 $\times$ 325 pixel$^{2}$ area was used for imaging polarimetry. The plate scale of the CCD is 1.48 arcsec pixel$^{-1}$ and the FOV is $\sim8\arcmin$. The full width at half maximum (FWHM) of the observed stellar image profile is found to be $\sim3$ pixels. The read-out noise and the gain of the CCD were 7.0 e$^{-1}$ and 11.98 e$^{-1}$ per Analog to Digital Unit, respectively. A R-band filter, matching the Kron-Cousin passband ($\lambda_{eff}$ = 0.760 $\mu$m), was used during the observations. AIMPOL provides only linear polarization and consists of an achromatic half wave plate (HWP) acting like a modulator and a Wollaston prism as a beam splitter. This set up provides two images (ordinary and extraordinary) of each target on the CCD frame. The HWP is rotated to obtain four normalized Stokes parameters, q[R(0$^{\circ}$)], u[R(22.5$^{\circ}$)], q1[R(45$^{\circ}$)] and u1[R(67.5$^{\circ}$)], corresponding to its four positions, i.e. 0$^{\circ}$, 22.5$^{\circ}$, 45$^{\circ}$ and 67.5$^{\circ}$. We estimate the errors in normalized Stokes parameters $\sigma_{R}$($\alpha$)($\sigma_{q}$, $\sigma_{u}$, $\sigma_{q1}$, $\sigma_{u1}$) in per cent using the relation given by \cite{1998A&AS..128..369R}. The average background counts have also been estimated around ordinary and extraordinary images of each star \citep{1998A&AS..128..369R}.

The contribution of instrumental polarization from the measurements was removed by observing a number of unpolarized standard stars from \cite{1992AJ....104.1563S}. We also observed six polarized standard stars (HD 236633, BD+59$^{\circ}$389, HD 19820, HD 204827, HD 25443, HD 15445) from the list given by \cite{1992AJ....104.1563S} to determine the reference direction of the polarizer during our observing runs. The measurements were used to obtain the zero-point offset with respect to the north, which was later applied to the position angles of the observed stars. After bias subtraction, flat correction of the images using the flux normalization formula from \cite{1998A&AS..128..369R}, we aligned and combined multiple images of a given field. The selection of the ordinary and extraordinary pair of each star from a given field was automated using a program written in the Python language. We performed photometry of the selected pairs using the Image Reduction and Analysis Facility (IRAF) DAOPHOT package to obtain the P\% and position angle ($\theta_{P}$) of each star. The details of the instrument used and data reduction procedure are given in \cite{2013MNRAS.432.1502S,2015A&A...573A..34S,2017MNRAS.465..559S}.

\subsection{Spectroscopic Observations}

We carried out spectroscopic observations of 4 of the 20 newly identified co-moving sources discussed in this work using the Hanle Faint Object Spectrograph Camera (HFOSC) mounted on the 2-m Himalayan Chandra Telescope (HCT) of the Indian Astronomical Observatory (IAO). We selected sources based on their relative brightness. The HFOSC is equipped with a 2k $\times$ 4k SITe CCD chip. We used a combination of slit 167l (slit width 1.92 arcsec $\times$ 11 arcmin) and grism 7, covering the wavelength range 3500-7500 \AA. This slit and grism combination gives a spectral resolution of $\Delta\lambda\sim$ 8 \AA. All the pre-processing and data reduction were performed in the standard manner using various tasks available with the IRAF. We used \textquote{splot} task to obtain the equivalent width of our observed sources. The log of the observations is given in Table \ref{tab:log_table}.

\begin{table}
	\centering
	\caption{Log of observations.}
	\label{tab:log_table}
	\begin{tabular}{p{0.4cm}p{6.6cm}}\hline
		Year & Month (Date)\\\hline\hline
        \multicolumn{2}{l}{Polarimetric Observations}\\
		2015 & Oct (11), Nov (2, 3, 15, 16, 17), Dec (15)\\
        2016 &	Oct (23, 25, 26, 27, 28), Nov (22, 26, 27)\\
        2017 & May (22, 23), Oct (13, 14, 17, 18, 19, 20, 21, 26, 27)\\
        \hline
        \multicolumn{2}{l}{Spectroscopic Observations}\\
        2018 & Oct (11), Nov (30)\\
        \hline     
	\end{tabular}\\
\end{table}

\subsection{\textit{Spitzer} photometry of point sources}

The \textit{Spitzer} photometry of the known YSO candidates are obtained from the literature \citep{2009ApJS..185..198K}. For the newly identified comoving sources, we calculated the \textit{Spitzer} IRAC (3.6, 4.5, 5.8, 8.0 $\mu$m) magnitudes by carrying out photometry on the images obtained from the \textit{Spitzer} Space Observatory archive with Program ID: 30574 \citep{2009ApJS..185..198K}. The basic calibrated data (BCD) frames were co-added and mosaic images were created using the SSC mosaicing and point-source extraction software \citep[MOPEX; ][]{2005PASP..117.1113M}. Sources were extracted from the final images using MOPEX. Photometry was performed by extracting the flux using an aperture of 7 pixels wide box centered on each source using the APEX tool developed by the \textit{Spitzer} Science Center. For the sources for which APEX failed to detect in automated mode at one or more wavelengths, we used the user list option to supply the coordinates of the source to obtain the flux values. We adopted 280.9, 179.7, 115.0 and 64.1 Jy in the 3.6, 4.5, 5.8 and 8.0 $\mu$m bands, respectively as the zero points for conversion between flux densities to the magnitudes as provided by \cite{2009ApJS..185..198K}. 

\subsection{The {\it Gaia} DR2 data}
 
\textit{Gaia} DR2 provides accurate positions, parallaxes and proper motions for more than a billion objects \citep{2018A&A...616A...1G,2018A&A...616A..10G}. However, the conversion from parallax to distance is known to become non-trivial when the observed parallax is small compared to its uncertainty especially in cases where $\sigma_{\varpi}$/$\varpi\gtrsim20\%$ \citep{2015PASP..127..994B}. Recently, \cite{2018AJ....156...58B} calculated distances to 1.331 billion sources for which \textit{Gaia} measured parallaxes by adopting an exponentially decreasing space density prior in distance. The distances to the YSO candidates and the field stars for which we made polarization measurements are obtained from \cite{2018AJ....156...58B} and proper motions from \cite{2018A&A...616A...1G}.

%####################################################################################
\section{Results and Discussion}\label{sec:res_dis}
\subsection{Distance to L1172/1174}

%####################################################################################
\subsubsection{Distance based on the known YSO candidates associated with the cloud}

One of the direct ways of estimating distances to a molecular cloud is to use the stars that are associated with the cloud. Recently, using 47 YSO candidates identified by \citet{2008hsf1.book..136K},  \citet{2018ApJ...867..151D} estimated a distance of 358$\pm$32 pc to the whole Cepheus flare region. But as discussed by \cite{1998ApJS..115...59K}, the Cepheus flare region contains clouds that are located at different distances in projection. The distances of YSOs studied by \citet{2018ApJ...867..151D} also show a large spread ranging from $\sim200$ pc to $\sim400$ pc. Therefore, in this work, we restricted our analysis to sources that are located within a $1\degr\times1\degr$ region about the star HD 200775 and estimated the distance to L1174. 

\begin{figure}
	\includegraphics[width=8.3cm, height=6.5cm]{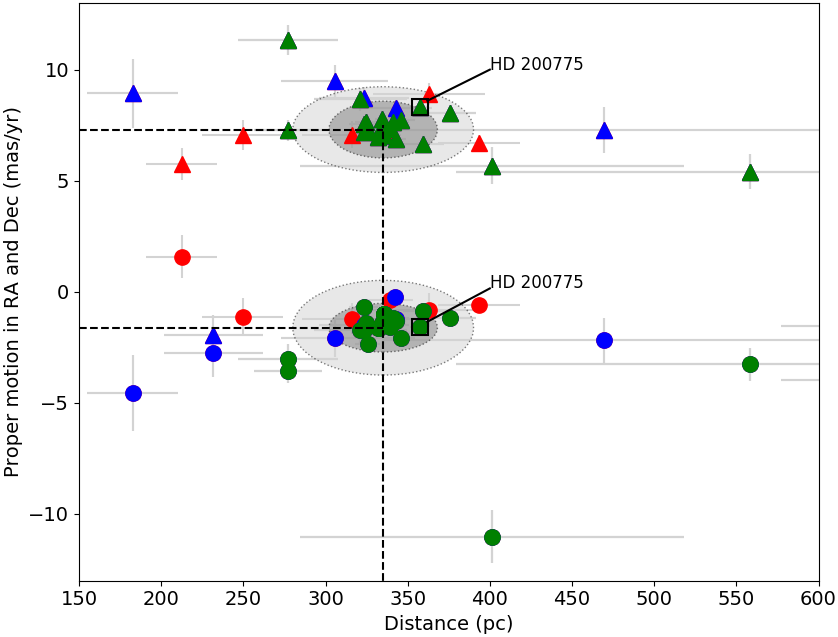}
	\caption{Proper motion values of the known YSO candidates associated with L1174 are plotted as a function of their distances obtained from \textit{Gaia} DR2. The triangles and circles represent the distance-$\mu_{\alpha\star}$ and distance-$\mu_{\delta}$ values respectively. Location of HD 200775 is also marked. The error ellipses corresponding to 3 $\times$ MAD (darker shade) and 5 $\times$ MAD (lighter shade) in proper motion and distance values are drawn. The dashed lines show the median values of distance, $\mu_{\alpha\star}$ and $\mu_{\delta}$. The ratios of distances and proper motion values with their respective errors 2>(m/$\sigma$m)$\geqslant$1 are for red points, 3>(m/$\sigma$m)$\geqslant$2 for blue points and (m/$\sigma$m)$\geqslant$3 for green points.}
	\label{fig:pm_YSO}
\end{figure}

A total of 58 YSO candidates have been identified till now (after taking into account the common sources among various studies) in the vicinity of L1172/1174 \citep{2009ApJS..185..198K, 2009ApJS..185..451K, 2013MNRAS.429..954Y}. Of the 58 YSO candidates, we found distance and proper motion values in the right ascension ($\mu_{\alpha\star}$=$\mu_{\alpha}$cos$\delta$) and declination ($\mu_{\delta}$) for 37 sources in the \citet{2018AJ....156...58B} and in the \textit{Gaia} DR2 \citep{2018A&A...616A...1G} catalogues, respectively. For all the 37 sources, a \textit{Gaia} counterpart was found well within 1$\arcsec$ of the input coordinate of the YSO candidates and the ratio, m/$\sigma$m (here m represents the distance, $\mu_{\alpha\star}$ and $\mu_{\delta}$ values and the $\sigma_{m}$ represents their respective errors) $\geqslant1$. The \textit{Gaia} DR2 results for the YSO candidates are shown in Fig. \ref{fig:pm_YSO}. The triangles and circles represent the distance-$\mu_{\alpha\star}$ and distance-$\mu_{\delta}$ values respectively. Of the 37, 6 sources (red circles and triangles) are with 1$\leqslant$m/$\sigma_{m}<2$; 8 sources (blue circles and triangles) are with $2\leqslant$m/$\sigma_{m}<3$; and 23 sources (green circles and triangles) with m/$\sigma_{m}\geqslant$3. The distance-$\mu_{\alpha\star}$ and distance-$\mu_{\delta}$ values for the HD 200775 are identified and marked separately. Three sources out of these 23 are found to be at a distance greater than 1 kpc. Thus, in this work, we selected only those sources for which m/$\sigma_{m}\geqslant$3 and are at distances less than 1 kpc. The results of these sources (20) are shown in Table \ref{tab:YSO_gaia}. 

Based on the proper motion and the distance values, a clear clustering of sources are evident in Fig. \ref{fig:pm_YSO}. The median of $\mu_{\alpha\star}$, $\mu_{\delta}$ and distance values are 7.301 mas/yr, -1.619 mas/yr and 335 pc, respectively. The variance and the standard deviation that are commonly used to measure spread in a data are more affected by the extreme (high and low) values. Therefore here we used median absolute deviation (MAD) to estimate statistical dispersion which is more resilient to outliers in a data set than the standard deviation. The estimated MAD in distance, $\mu_{\alpha\star}$ and $\mu_{\delta}$ are $11$ pc, $0.386$ mas/yr and $0.427$ mas/yr, respectively. A majority of the YSO candidates (14) are found to fall within the constraints of three times the MAD in proper motions and distances as shown in Fig \ref{fig:pm_YSO} using the ellipses drawn with darker shade. Two more sources get included when we consider a constraint of five times the MAD in proper motions and distances (ellipses drawn in lighter shade). Remaining four sources show large scatter from the median values. We note that HD 200775 is showing a larger distance ($\sim2 \times$ MAD) with respect to the median value. In column 9 of Table \ref{tab:YSO_gaia}, we give the renormalized unit weight error \citep[RUWE\footnote{RUWE values are obtained from \url{http://gaia.ari.uni-heidelberg.de/}};][]{LL:LL-124} for the YSO candidates. RUWE is a scaled unit weight error (UWE) with the scaling factor depending on the magnitude and color of the sources. Theoretically, the UWE is expected to be close to 1.0 for well-fitted solutions of single stars but can show larger values depending on the source environment and the geometric properties such as binarity. Practically, sources showing RUWE $\leq 1.4$ are considered as having good astrometric solution \citep{LL:LL-124}. As can be noticed, of the 20 YSO candidates, 10 sources show RUWE$>$1.4. The four sources showing large scatter from the median values are the ones showing relatively high values of RUWE ($>$3.0). Of the remaining six, except one, the others show RUWE in the range of 1.5-1.6, including HD 200775 which shows a marginally higher value of 1.6. Thus we considered all the 16 sources that are lying within 5 $\times$ MAD limit as part of L1174 and sources (four of them) that are located outside of the 5 $\times$ MAD ellipses are considered as outliers. The distance of 335$\pm$11 pc (the MAD is taken as the uncertainty in the distance) to L1174 implies that L1174 is at a height of $\sim80$ pc above the galactic mid-plane. This is higher when compared to the scale-heights of $\sim45$ pc estimated for the OB stars in the local galactic disc \citep{2000AJ....120..314R} and for the molecular gas \citep[e.g.][]{2015ARA&A..53..583H}. Thus, L1174 is an example of a sparse cluster of young stars (which contains one or two intermediate mass stars) forming at a relatively high galactic latitude.

%************************************************

%***************************************************************
\begin{figure*}
\includegraphics[height=9.5cm, width=\textwidth]{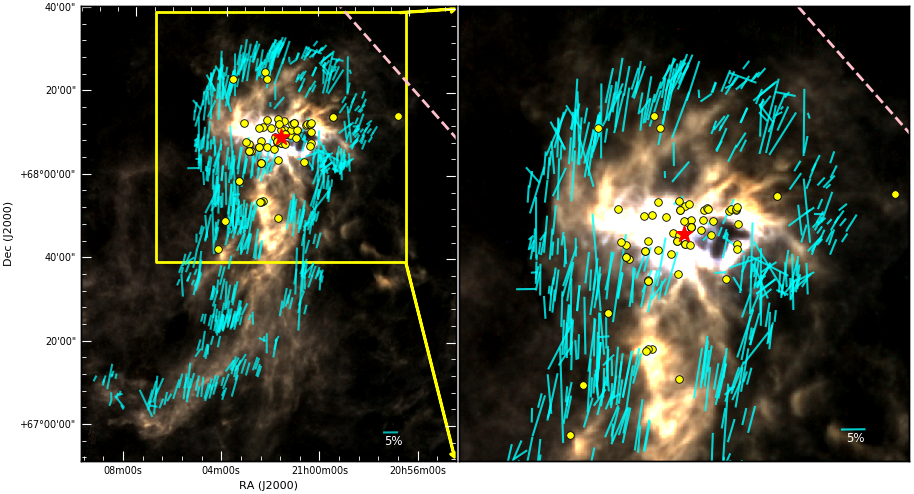}
\caption{{\bf LHS:} R-band polarization vectors (cyan lines), obtained from our observations, overplotted on {\it Herschel} colour-composite diagram with 250, 350 and 500 $\mu$m images. Broken line in pink represents the galactic plane. Locations of the known YSO candidates (yellow circles) around the central star HD 200775 (red star) are also shown. A polarization vector corresponding to 5\% is shown for reference. {\bf RHS:} Enlarged view of the  1$^{\circ}\times$1$^{\circ}$ region around HD 200775. Symbols represent the same as LHS.}
    \label{fig:NGC7023_CC_pol_vecs}
\end{figure*}
%***************************************************************

%####################################################################################
\subsubsection{R-band polarization and the {\it Gaia} DR2 parallax measurements of field stars}

Presence of interstellar dust grains along a given line of sight can be inferred by their effects on starlight coming from the background stars. Rotating non-spherical dust grains get aligned with their minor axis parallel to the ambient magnetic field \citep{1951ApJ...114..206D, 2012ARA&A..50...29C}. When the unpolarized starlight coming from a background star passes through regions containing such dust grains, the light gets polarized due to the selective absorption. Normally the value of P\% increases gradually with the increase in the column of dust grains along the pencil-beam of a starlight. However, when it encounters a molecular cloud, a sudden increase in the values of P\% occurs. Therefore, while the stars that are foreground to the cloud are expected to show low values of P\%, the stars that are located behind the cloud are expected to show higher values of P\%. The distance at which the sudden increase in the P\% occurs is taken as the distance of the cloud \citep[e.g.][]{1992BaltA...1..149S,1997A&A...327.1194W,1998A&A...338..897K,2007A&A...470..597A}. We examined the R-band polarization measurements of 569 sources that are projected on L1172/1174. The advantage of using polarization measurements to infer the presence of a molecular cloud is that the measured values are independent of the nature of the background stars. 

\begin{figure}
    \includegraphics[width=8.3cm, height=13cm]{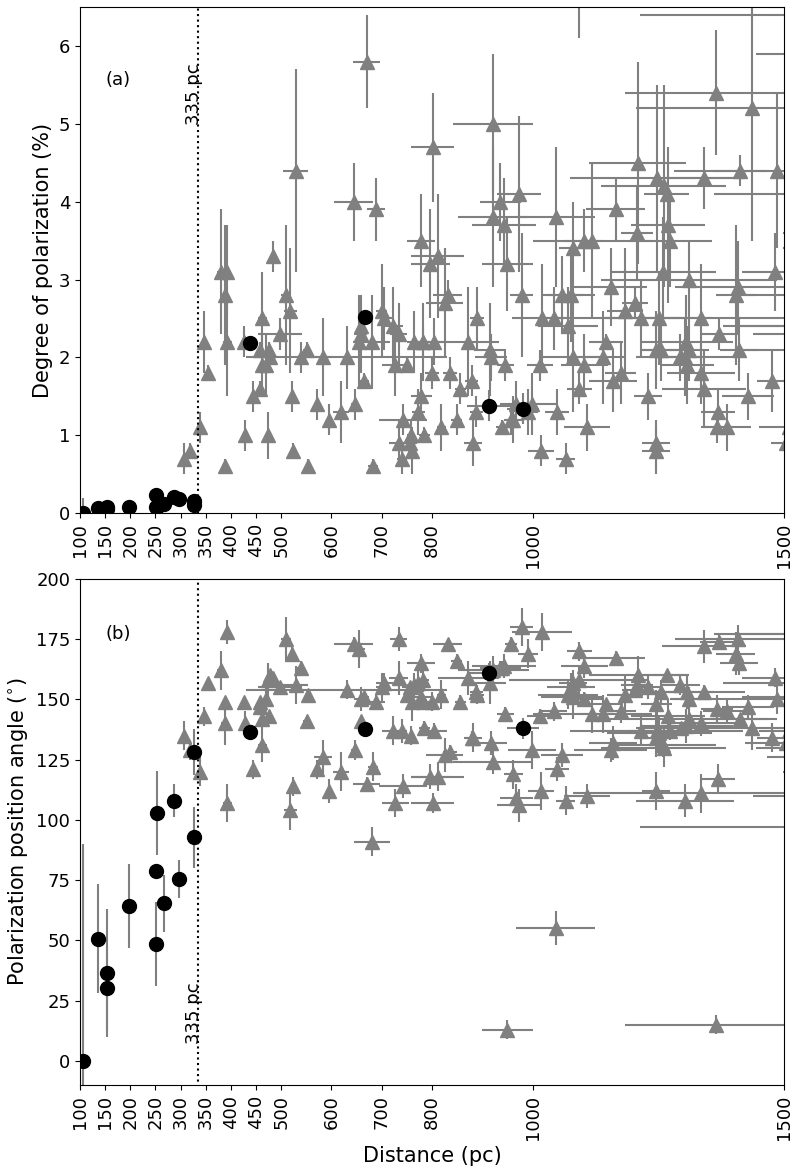}
    \caption{{\bf (a)} Polarization (\%) vs. distance plot for the stars projected towards the direction of L1172/1174 (filled triangles in grey). The filled circles in black are the sources for which the P\% is obtained from the \citet{2000AJ....119..923H} catalogue. {\bf (b)} Polarization position angle vs. distance plot for the stars projected in the periphery of L1172/1174. The symbols are same as above.} \label{fig:pol_figure}
\end{figure}

The results of our polarization measurements for the stars are plotted in Fig. \ref{fig:NGC7023_CC_pol_vecs}. The lengths of the vectors are proportional to the value of P\% and the orientations depend on the $\theta_{P}$. The values of $\theta_{P}$ are measured from the north increasing towards the east. We obtained \textit{Gaia} DR2 distances for 545 sources from the \citet{2018AJ....156...58B} catalogue. In all the cases, we obtained a \textit{Gaia} DR2 counterpart within 1$\arcsec$ of our source positions. The results are presented in Fig. \ref{fig:pol_figure}. The P\% vs. distance and $\theta_{P}$ vs. distance plots are shown in Fig. \ref{fig:pol_figure} (a) and (b), respectively using filled triangles in grey. We have shown only 301 sources for which the ratios of the P\%, $\theta_{P}$ and distance and their corresponding error, P/$\sigma_{P}$, $\theta_{P}$/$\sigma_{\theta_{P}}$ and $d$/$\sigma_{d}$ $\geqslant3$, respectively. The mean values of P\% and $\theta_{P}$ are found to be 2.5\% and 142$^{\degr}$. The closest star observed by us is at a distance of 306 pc. Therefore to investigate the foreground contribution to the polarization, we searched for additional sources in the \citet{2000AJ....119..923H} catalogue with a search radius of 6$\degr$ about HD 200775. The search gave a total of 26 sources of which HD 200775, HD 203467, HD 208947 and HD 193533 were not considered in the analysis due to the following reasons. The polarization values of HD 200775 could be affected by the reflection nebulously around it. The HD 203467 is a Be type object, HD 208947 is an Algol variable and HD 193533 is classified as a variable source in the Simbad database. Of the 22, we obtained the \textit{Gaia} DR2 distance for 17 sources from the \citet{2018AJ....156...58B} catalogue. The degree of polarization (P$_{H}$\%) and polarization position angles ($\theta_{H}$) of the 17 sources are shown using filled circles in black in both (a) and (b) of Fig. \ref{fig:pol_figure} respectively. Because the foreground sources show relatively very low polarization, we set no constraints on the P$_{H}$/$\sigma_{H}$ values while selecting them.

%**************************************************************
%**************************************************************

In Fig. \ref{fig:pol_figure} (a), the P$_{H}$\% of the stars having distances less than $\sim335$ pc (marked using dotted line in both upper and lower panels) show low values of polarization while the ones behind the cloud show relatively higher values. A significant increase in the P\% values are seen at $\sim335$ pc for the sources observed by us which confirms the presence of the cloud at that distance, similar to what we obtained from the YSO candidates. From the Fig. \ref{fig:pol_figure} (b), as the distance increases from $\sim100$ pc to $\sim335$ pc, the $\theta_{H}$ values are found to change systematically from $\sim0\degr$ to a value close to the mean value of 142$^{\degr}$ obtained for the sources observed by us. The mean value of the $\theta_{H}$ for the 4 sources lying beyond the distance of 335 pc is found to be $\sim143\degr$. Based on an abrupt increase in the values of the P\% and the change in the $\theta_{P}$ for stars projected on the cloud, we confirm the distance of the entire L1172/1174 cloud complex at $\sim335$ pc. 

\subsection{Additional comoving sources identified from the \textit{Gaia} DR2 proper motion}

With the knowledge of the precise values of position, proper motion and distance from the \textit{Gaia} DR2 measurements, we looked for additional, probably young, sources that are comoving with the already known YSO candidates associated with L1172/1174. For this, we obtained proper motions and distances of sources that are located within a region of 1$^{\circ}\times1^{\circ}$ centered around HD 200775 from the \textit{Gaia} DR2 and \citet{2018AJ....156...58B} catalogues, respectively. For the reason that the majority of the known YSO candidates are distributed within a region of 1$^{\circ}\times1^{\circ}$ centered around HD 200775, we conducted our search for the comoving objects also within the same area. As in the case of the YSO candidates, here again we selected only those sources for which m/$\sigma$m$\geqslant3$. Considering that a majority of the previously known YSO candidates are found within 5 $\times$ MAD with respect to the median values of the distance and the proper motions, the same criteria are used to select the comoving sources also. That is, all sources falling within the range of 5 $\times$MAD with respect to the median values of the distance and the proper motions are regarded as comoving sources and included for further analysis.

%********************************************************************************
\begin{figure}
    \includegraphics[height=7cm, width=8.3cm]{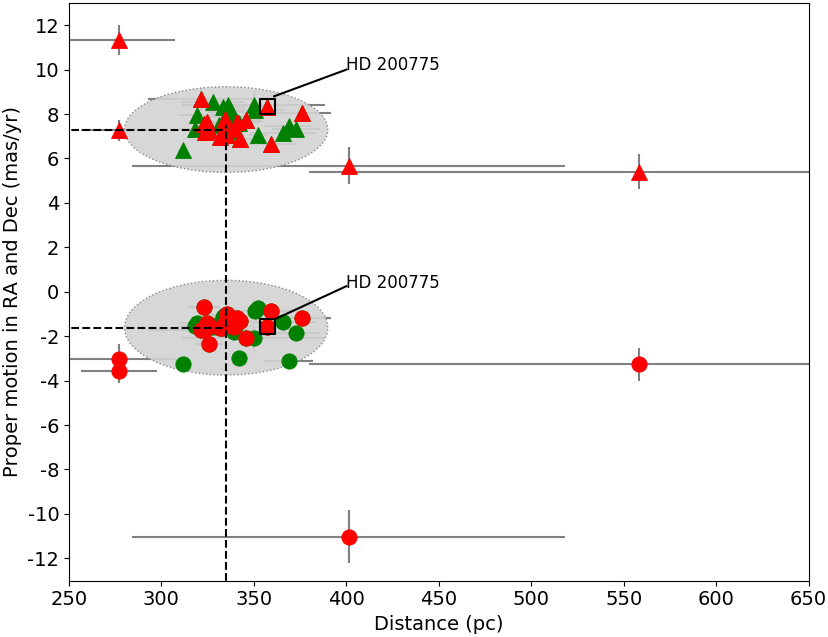}	
    \caption{Proper motion vs. distance plot for the known YSO candidates (triangles and circles in red) and for the 20 newly identified comoving sources (triangles and circles in green). The triangles (red and green) and circles (red and green) represent the distance-$\mu_{\alpha\star}$ and distance-$\mu_{\delta}$ values of the known YSO candidates and comoving sources respectively. Locations of HD 200775 are identified by square boxes. The grey ellipses represent the boundary of the proper motion values and the distance ranges used to identify the new comoving sources. The dashed lines show the median values of distance, $\mu_{\alpha\star}$ and $\mu_{\delta}$.} 
    \label{fig:prop_motn_figure}
\end{figure}
%********************************************************************************

\begin{figure}
    \includegraphics[height=7.5cm, width=8.3cm]{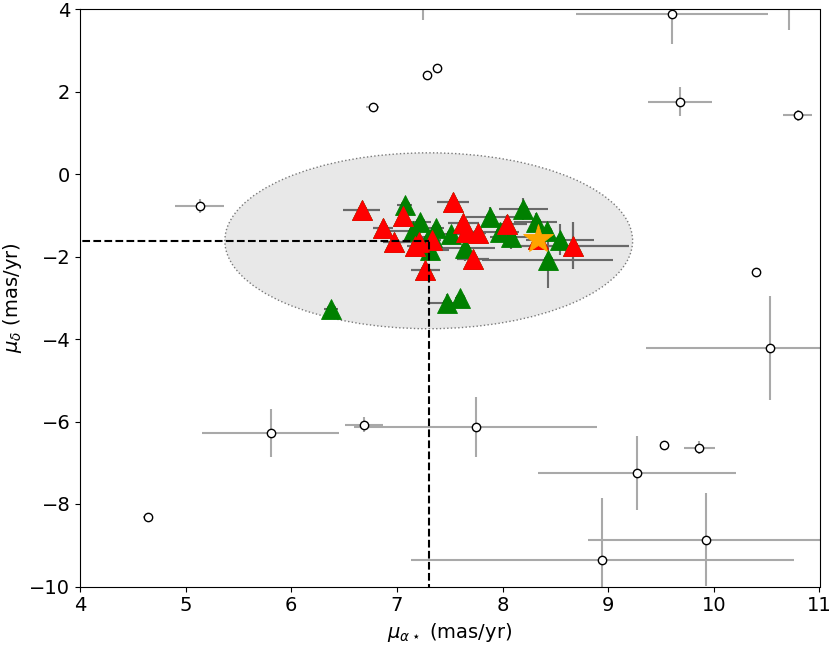}	
    \caption{$\mu_{\delta}$ vs. $\mu_{\alpha\star}$ plot for the known YSO candidates and for the comoving sources. Red triangles represent proper motion values of known YSO candidates and green triangles represent that of the comoving sources. The grey ellipse represents the boundary of the proper motion values considered to select the comoving sources. The orange star symbol indicates the location of HD 200775. The open circles represent sources not satisfying the 5 $\times$ MAD conditions in distance and proper motion values. The dashed lines show the median values of $\mu_{\alpha\star}$ and $\mu_{\delta}$.} 
    \label{fig:prop_motn_radec}
\end{figure}

\begin{figure}
    \includegraphics[height=7.5cm, width=8.3cm]{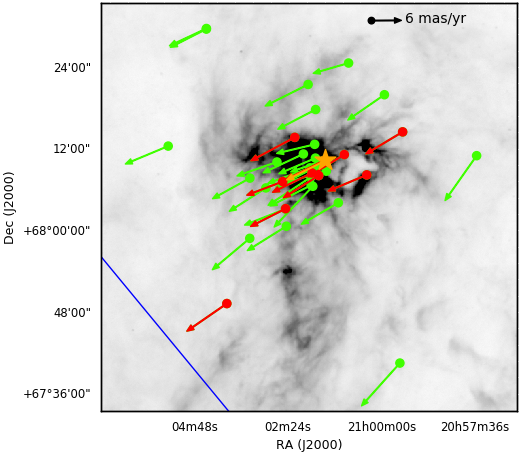}
    \caption{Proper motion plot for the YSO candidates (red arrows) associated with L1174 overplotted on the \textit{Herschel} SPIRE 250 $\mu$m image. The green arrows represent the same for the newly identified comoving sources. HD 200775 is indicated by a star symbol and an arrow in orange. The blue solid line represents the galactic plane.} 
    \label{fig:prop_motn_YSO}
\end{figure}

The selected sources are shown on the proper motion $-$ distance plot presented in Fig. \ref{fig:prop_motn_figure} and results are tabulated in Table \ref{tab:YSO_new}. The distance-$\mu_{\alpha\star}$ and distance- $\mu_{\delta}$ values of the known YSO candidates are shown using filled triangles and filled circles in red, respectively. The same for the comoving sources are presented using filled triangles and filled circles in green, respectively. The ellipses drawn in grey in Fig. \ref{fig:prop_motn_figure} represent the constraint of 5 $\times$ MAD drawn at the median values of $\mu_{\alpha\star}$, $\mu_{\delta}$ and distances of the known YSO candidates. Here again, the distance-$\mu_{\alpha\star}$ and distance-$\mu_{\delta}$ values of HD 200775 are marked. We found a total of 20 additional sources that are located within the grey ellipses. The distance-$\mu_{\alpha\star}$ and distance-$\mu_{\delta}$ values of these 20 additional sources are shown using filled triangles and filled circles in green respectively in Fig. \ref{fig:prop_motn_figure}. In Fig. \ref{fig:prop_motn_radec} we show the $\mu_{\alpha\star}$ and $\mu_{\delta}$ values of the known YSO candidates and the 20 additional sources using filled triangles in red and green respectively. HD 200775 is indicated by an orange star symbol. The grey ellipse indicates the range of 5 $\times$ MAD in $\mu_{\alpha\star}$ and $\mu_{\delta}$ about the median values obtained for the known YSO candidates. The kinematic association of the newly found 20 sources with the known YSO candidates is clearly evident in Fig. \ref{fig:prop_motn_figure} and Fig. \ref{fig:prop_motn_radec}. The proper motion values of other sources (open circles) in the field are found to be very different. The positions of the known YSO candidates and the 20 additional sources that are comoving are identified on the \textit{Herschel} SPIRE 250 $\mu$m image as shown in Fig. \ref{fig:prop_motn_YSO} using red and green circles, respectively. The position of HD 200775 is shown using an orange star symbol. The arrows in red, green and orange show the proper motion directions of the known YSO candidates, newly identified comoving sources and HD 200775, respectively. Spatial locations of the comoving sources are very similar to those of the known YSO candidates. The distance and the proper motion values of the comoving sources are listed in Table~\ref{tab:YSO_new}. The RUWE value for the comoving sources are also shown in Table ~\ref{tab:YSO_new}. Of the 20 comoving sources, only one source (c12) show RUWE of 1.5 which is marginally greater than the value of 1.4 considered as criterion for good solutions.

%********************************************************************************
%********************************************************************************
%********************************************************************************

To investigate whether the sources with similar proper motions as that of the known YSO candidates from a given 1\degr~ square area are a chance projection of field stars or are really the comoving sources associated with the region, we selected 1\degr square area from three reference fields lying towards the direction of the southern (tail) part of L1172/1174 having similar declination but differing in right ascension (field 1: 318.79282$^{\circ}$, 67.174828$^{\circ}$; field 2: 316. 21048$^{\circ}$, 67.206136$^{\circ}$; field 3: 313.56081$^{\circ}$, 67.197843$^{\circ}$). Here also, we selected only those sources that are with m/$\sigma_{m}\geqslant$3. We found only one source (in field 1), HD 202461 (distance = 309$^{3}_{-2}$ pc, $\mu_{\alpha\star}$ = 7.317$\pm$0.060 mas/yr and $\mu_{\delta}$ = -1.724$\pm$0.050 mas/yr), which satisfies the above constraints (marked with a square box in the ellipses). Information pertaining to this source is not available in the literature. Thus, on an average, we expect at most one source in a given 1\degr square area in the direction close to L1172/1174 that could be a chance projection of field stars having proper motions and distances similar to those of the known YSO candidates. This confirms the presence of an over-density of stars within 1 square degree around HD 200775.

%#############################################################################
\subsection{Properties of the sources identified around HD 200775}

\subsubsection{X-ray properties}\label{sec:x_ray}

The field containing HD 200775 was observed by the \textit{XMM-Newton} telescope to investigate X-ray properties of early type stars \citep{2014ApJS..215...10N}. We searched for additional X-ray sources in the region around HD 200775 in the XMM-SSC, 2018 catalogue \citep{2016A&A...590A...1R}. Within the FOV of the \textit{XMM-Newton}, we found a total of 35 X-ray emitting sources. The maximum likelihood parameter is set at $\geqslant15$ \citep[which gives the maximum likelihood of the source detection, ][]{2008A&A...491..961L}, and the detection quality flag set at 0 or 1. The exposure time of the observations was 11.9 ks. Out of 35, 19 sources are found to have the ratio of the flux in 0.2-12 keV and the corresponding error $\geqslant$ 3. Of these, 10 known YSO candidates (including HD 200775) and 4 newly identified comoving sources are found to spatially coincide within 5$^{\prime\prime}$ from the positions of the X-ray detection. Based on the study conducted by \citet{2011ApJS..194....3G} on the Carina Nebula using the Chandra telescope and scaling for the survey coverage, approximately $\sim10-20$ foreground and background and 30 extragalactic sources could be detected within the FOV of \textit{XMM-Newton}. The foreground and background stars may have detection in \textit{WISE}, but extragalactic sources may not have a \textit{WISE} counterpart, because they are below its sensitivity limit. Out of 10 known YSO candidates, 8 of them have \textit{WISE} counterparts. All the 4 newly identified comoving sources emitting in X-ray, have \textit{WISE} counterparts. 

\begin{figure}
    \includegraphics[height=7.5cm, width=8.3cm]{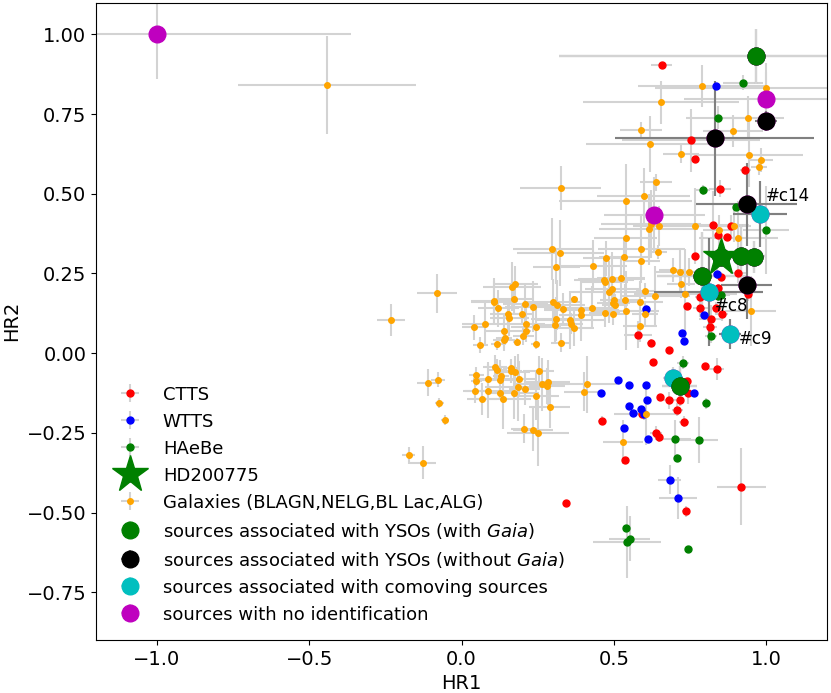}
    \caption{Hardness ratio plot for the X-ray sources detected by the \textit{XMM-Newton} telescope in the vicinity of HD 200775. Known YSO candidates (with and without reliable \textit{Gaia} data) and the newly identified comoving sources are identified. The comoving sources which are observed by us spectroscopically are marked. The hardness ratios of CTTS (red dots), WTTS (blue dots), HAeBe (green dots) and galaxies (orange dots) are also shown.}
    \label{fig:HR_xmm}
\end{figure}

The XMM-SSC catalogue provides hardness ratios, HR1 and HR2 also. These are X-ray colours defined as (H-S)/(H+S) where H and S for HR1 are flux values in the band 0.5-1.0 keV and 0.2-0.5 keV respectively. For HR2, the H and S are flux values in the bands 1.0-2.0 keV and 0.5-1.0 keV respectively. Fig. \ref{fig:HR_xmm} shows the HR2 vs HR1 plot for our YSO candidates (filled circles in green and black), 4 newly identified comoving sources (filled circles in cyan) and HD 200775 (star symbol in green). We obtained \textit{XMM-Newton} data for the CTTS and WTTS taken from \citet{2010ApJ...724..835W} which are shown using red and blue filled dots respectively. Herbig AeBe (HAeBe) sources taken from \citet{1994A&AS..104..315T} catalogue are shown using dots in green. As mentioned already, some of the X-ray detection found in the vicinity of HD 200775 could be background galaxies. To differentiate the background galaxies from the PMS stars based on their hardness ratios, we obtained \textit{XMM-Newton} data around a 5$^{\prime\prime}$ search radius of the galaxy samples of \cite{2007A&A...476.1191B}. This includes broad line active galactic nuclei, narrow emission line galaxies, absorption line galaxies, BL Lacertae. These sources are shown using orange dots in Fig. \ref{fig:HR_xmm}.

%*********************************************************************************
%*********************************************************************************
The HAeBe, CTTS and WTTS \citep{1994A&AS..104..315T,2010ApJ...724..835W} are found to occupy a distinct region in Fig. \ref{fig:HR_xmm}. However, the distribution of the galaxy sample is noticeably different from that of the HAeBe, CTTS and WTTS. The HR1 colours of CTTS range between 0.5 - 1.0, while for WTTS, it is mostly close to $\sim0.6$. The HR2 colours for CTTS show large range (0.5 to -0.5), while WTTS predominantly show HR2 $\lesssim0.5$. The known YSO candidates and the comoving sources found associated with the X-ray detection are distributed in a similar manner as that of HAeBe, CTTS and WTTS. Here our aim is not to characterize their X-ray properties but to show that X-ray sources found in the vicinity of HD 200775 show X-ray colours similar to those of CTTS and WTTS. Of the 5 X-ray detections that are not associated with any of the known YSO candidates or the comoving sources, 2 of them, namely, \#x1 and \#x5 (the identification numbers are as given in Table \ref{tab:xray_table}) do not have 2MASS, \textit{Spitzer} and \textit{WISE} counterparts within $5^{\prime\prime}$ search radius. The location of \#x1 at (-1.0, 1.0) is conspicuously different from the rest of the sources in Fig. \ref{fig:HR_xmm} while the source \#x5, located at (1.0, 0.8), could possibly be an extragalactic source due to the lack of \textit{WISE} detection. Among the 3 sources with \textit{WISE} counterparts, \#x2 which is located at (0.63, 0.43) in Fig. \ref{fig:HR_xmm} shows hardness ratios consistent with those of the extragalactic sources. The sources \#x3 and \#x4, based on the \textit{Gaia} DR2 distances, are foreground sources and hence not shown in the figure. Two \textit{Gaia} counterparts are found within our search radius of 5\arcsec for \#x3. The high proper motions ($\mu_{\alpha\star}$= -2.492 mas/yr and -4.615 mas/yr and $\mu_{\delta}$= -34.948 mas/yr and -34.931 mas/yr) confirm that they are foreground. \#x4 with $\mu_{\alpha\star}$= 8.628 mas/yr and $\mu_{\delta}$= 5.798 mas/yr, confirms that this source is kinematically not associated with the region. Therefore we did not consider any of the X-ray sources other than those associated with the known YSO candidates and the comoving sources in further analysis. 

\subsubsection{Spectroscopy of four comoving sources}\label{sec:spec}

\begin{figure}
    \includegraphics[width=8.3cm, height=7.5cm]{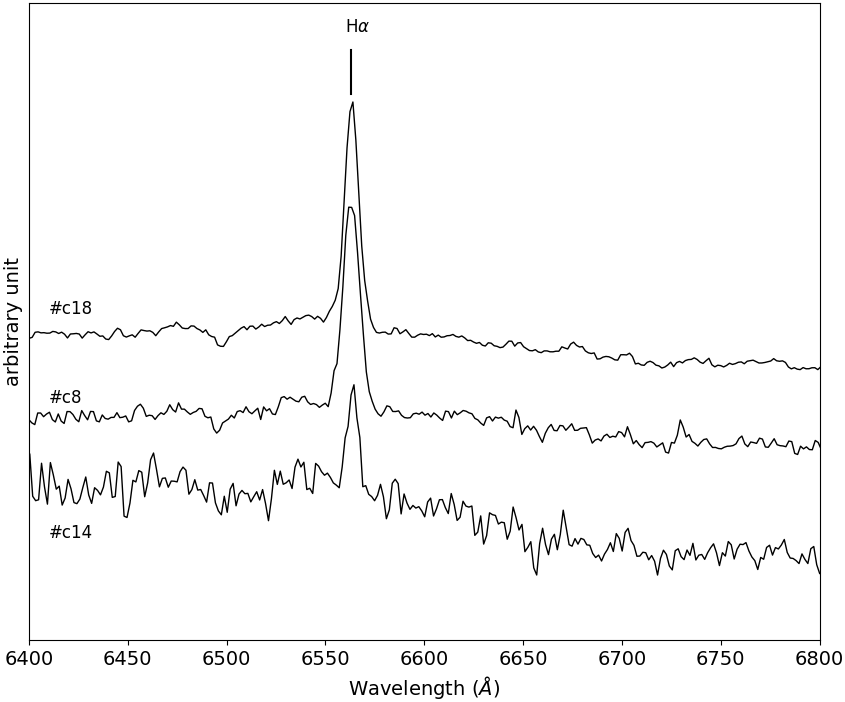}
    \caption{Spectra of 3 of the 4 newly identified comoving sources found around HD 200775. The source identification numbers are also given. Wavelengths corresponding to H${\alpha}$ lines are marked with dashed vertical lines.}
    \label{fig:com_spctrum}
\end{figure}

\begin{figure}
    \includegraphics[width=8.2cm, height=7.0cm]{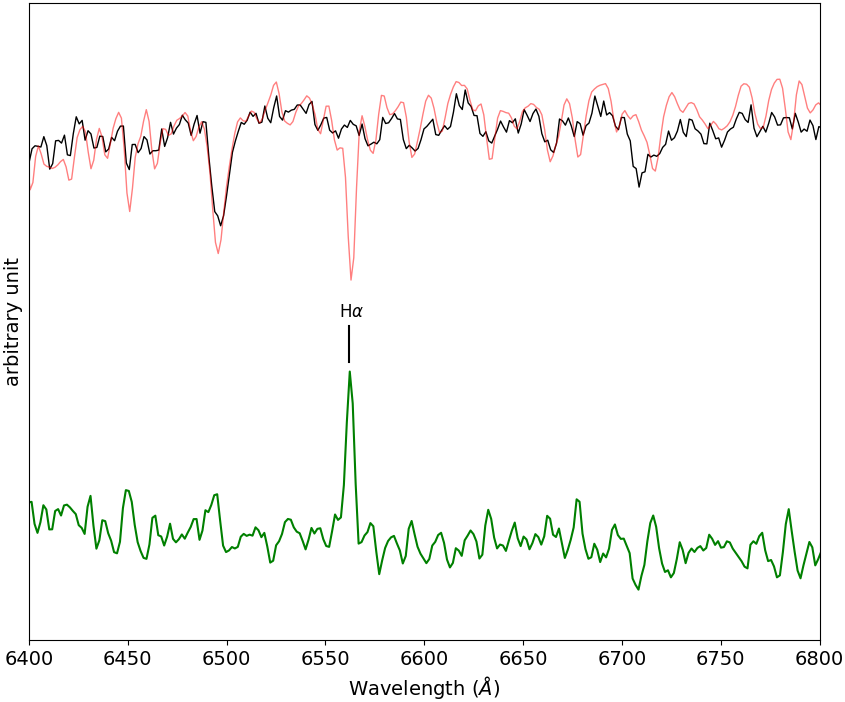}
    \caption{Spectrum of the comoving source \#c9 (black) overplotted with a K2 spectral type spectrum (red) from \citet{1984ApJS...56..257J}. The subtracted spectrum (\#c9-K2) shown in green reveals the filled-in emission in the observed spectrum of \#c9.}
    \label{fig:com_resultant}
\end{figure}

Of the 20 newly identified comoving sources, we obtained spectroscopic observations for four of them during our recent observing run. The objective was, as a first step, to look for emission lines in them. Spectral types of the four sources were determined by comparing the features in the spectrum of our sources with those in the templates of main sequence stars from the stellar library provided by \citet{1984ApJS...56..257J}. Before performing the comparison, we normalized the spectra of our sources and took the templates to a common resolution. The uncertainty in our spectral classification is found to be of two spectral subclasses. The stars \#c8 and \#c18 (star identification numbers are as given in Table \ref{tab:YSO_new}) are found to be of M1 spectral type while \#c14 is found to be of M3 spectral type. The observed spectra of star \#c8, \#c14 and \#c18 are shown in Fig. \ref{fig:com_spctrum}. All the three are found to be X-ray emitters. We detected H${\alpha}$ in emission in all three of them. We used the SPLOT task in the IRAF to obtain EW(H$\alpha$) of our observed sources. The EW(H$\alpha$) of \#c8 and \#c18 are found to be -15.03$\pm$1.52~\AA~and~-13.01$\pm$1.30 \AA,~respectively. According to \cite{2003AJ....126.2997B}, these two sources can be classified as CTTS. However, it is surprising that these sources were not detected in any of the earlier surveys for H${\alpha}$ emission stars. Variability of H$\alpha$ emission in \#c8 and \#c18 could be a plausible reason. The EW(H$\alpha$) of \#c14 is found to be -4.06$\pm$0.41 \AA ~which could possibly be a WTTS as mentioned in \cite{2003AJ....126.2997B}.~The fourth star, \#c9, is having H${\alpha}$ in filled-in emission. This star is found to be of a K2 spectral type. To reveal its H$\alpha$ emission, we have subtracted the spectrum of a main sequence K2 spectrum (red) from that of the source \#c9 (black) as shown in Fig. \ref{fig:com_resultant}. The resultant spectrum is shown in green. The EW(H$\alpha$) obtained from the resultant spectrum is found to be -0.69$\pm$0.08~\AA. 

%**************************************************************************
%**************************************************************************
%**************************************************************************
 
%####################################################################################

\subsubsection{NIR and Mid-IR properties}\label{sec:nir_mir}

\begin{figure}
    \includegraphics[width=8.3cm, height=7.5cm]{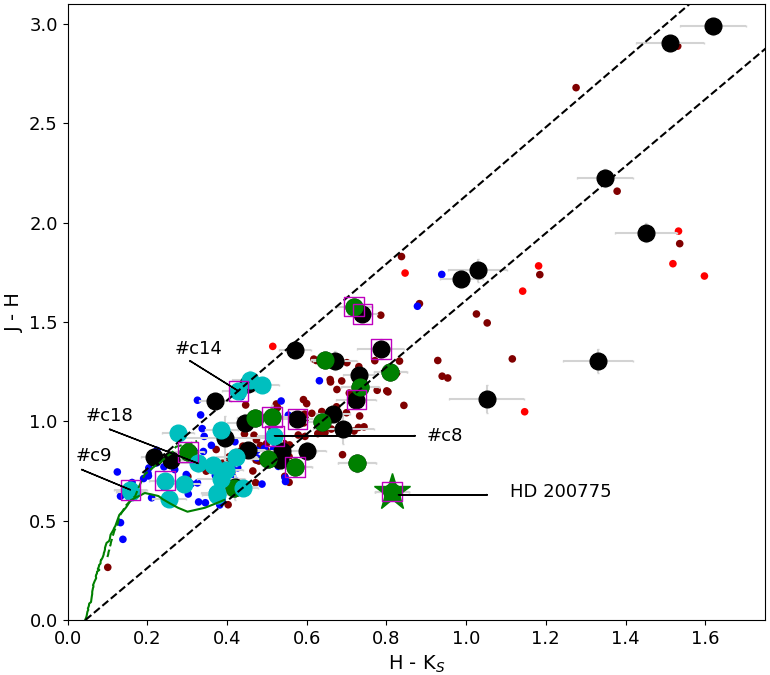}
    \caption{ The (\textit{J-H}) vs. (\textit{H-K$_{S}$}) CC diagram for the newly identified comoving sources shown using filled cyan circles. The known YSO candidates having reliable \textit{Gaia} data (green circle) and those without reliable \textit{Gaia} data or no detection (black circle) are also shown. The solid curves in green represent the locii of the unreddened main sequence stars and the giants. The Class I, II and III sources taken from \citet{2010ApJS..186..259R} are shown in red, maroon and blue dots, respectively. The known YSO candidates and the comoving sources spatially found to be associated with the \textit{XMM-Newton} X-ray detection are identified using square boxes in magenta. The 4 sources observed by us spectroscopically are identified and marked.}
    \label{fig:2mass_CC}
\end{figure}

To understand the nature of the newly identified comoving sources based on their near and mid-infrared colours, we obtained their 2MASS and \textit{Spitzer} magnitudes. The 2MASS magnitudes are obtained from \citet{2003yCat.2246....0C}. Sources having photometric quality `A' (SNR$\geqslant$10) in \textit{J}, \textit{H} and \textit{K$_{S}$} are selected. Out of 20 comoving sources, we found a 2MASS counterpart for 19 of them. Of the 20 known YSO candidates having reliable \textit{Gaia} DR2 data (Table \ref{tab:YSO_gaia}), we found a 2MASS counterpart for 17 of them. We used a search radius of $1^{\prime\prime}$ for getting the 2MASS counterparts. The comoving sources and the known YSO candidates are shown in the (\textit{J-H}) vs. (\textit{H-K$_{S}$}) colour-colour (CC) diagrams in Fig. \ref{fig:2mass_CC} using filled circles in cyan and green, respectively and listed in Table \ref{tab:2mass_wise_yso_com}. Also shown are the positions of Class I, II and III (red, maroon and blue dots, respectively) sources taken from \citet{2010ApJS..186..259R} for comparison. The four sources for which we made optical spectroscopic observations (discussed in section \ref{sec:spec}) are identified and labeled. Of the remaining 38 known YSO candidates lacking reliable data or detection by the \textit{Gaia}, we obtained the 2MASS magnitudes for 28 sources. The results are presented in Table \ref{tab:NRI_other} and shown in Fig. \ref{fig:2mass_CC} using filled circles in black. The known YSO candidates and the newly identified comoving sources that are spatially associated with the X-ray detection made by the \textit{XMM-Newton} satellite are identified using open square symbols in magenta. 

The distribution of the known YSO candidates in the (\textit{J-H}) vs. (\textit{H-K$_{S}$}) CC diagram as shown in Fig. \ref{fig:2mass_CC} is found to be similar to that of the Class I and Class II (red and maroon dots, respectively) sources taken from \citet{2010ApJS..186..259R}. A few of these known YSO candidates show relatively high amount of extinction. In contrast, the majority of the newly identified comoving sources are distributed to the left and lying below the known YSO candidates identified around HD 200775. This suggests that the comoving sources suffer relatively less extinction and exhibit small amount of near-IR excess emission compared to the known YSO candidates. A notable number of them fall in a space between the region occupied by the Class II and Class III (maroon and blue dots, respectively) sources \citep{2010ApJS..186..259R} which indicates that a significant number of them may have some amount of circumstellar material.

%********************************************************************************
%**************************************************************************************
\begin{figure}
    \includegraphics[width=8.3cm, height=7.5cm]{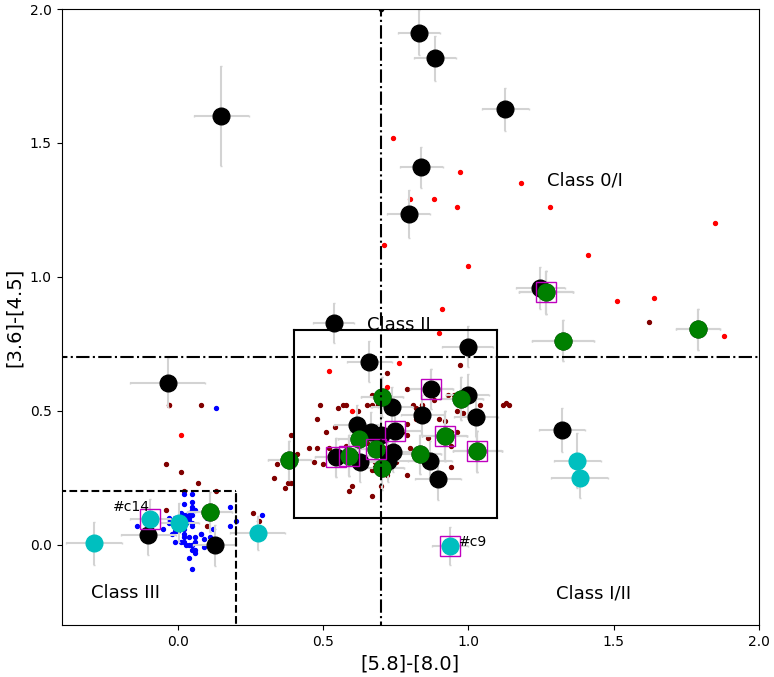}
    \caption{ The {\it Spitzer} CC diagram for the newly identified comoving sources shown using filled cyan circles. The known YSO candidates having reliable \textit{Gaia} data (green circle) and those without reliable \textit{Gaia} data or no detection by the \textit{Gaia} (black circle) are also shown. The Class I, II and III sources taken from \citet{2010ApJS..186..259R} are shown in red, maroon and blue dots, respectively. %Maroon open circles represent the TD sources taken from \citet{2012ApJ...750..157C}. 
    The known YSO candidates and the comoving sources spatially found to be associated with the \textit{XMM-Newton} X-ray detection are marked using square boxes in magenta. The 2 sources which are spectroscopically observed by us and have \textit{Spitzer} colours are marked. The boundaries within which Class I, II and III sources generally occupy \citet{2009A&A...504..461F,2010ApJ...717.1067C} are also shown.}
    \label{fig:spitzer_CC}
\end{figure}
%**************************************************************************************
\begin{figure}
    \includegraphics[width=8.3cm, height=7.5cm]{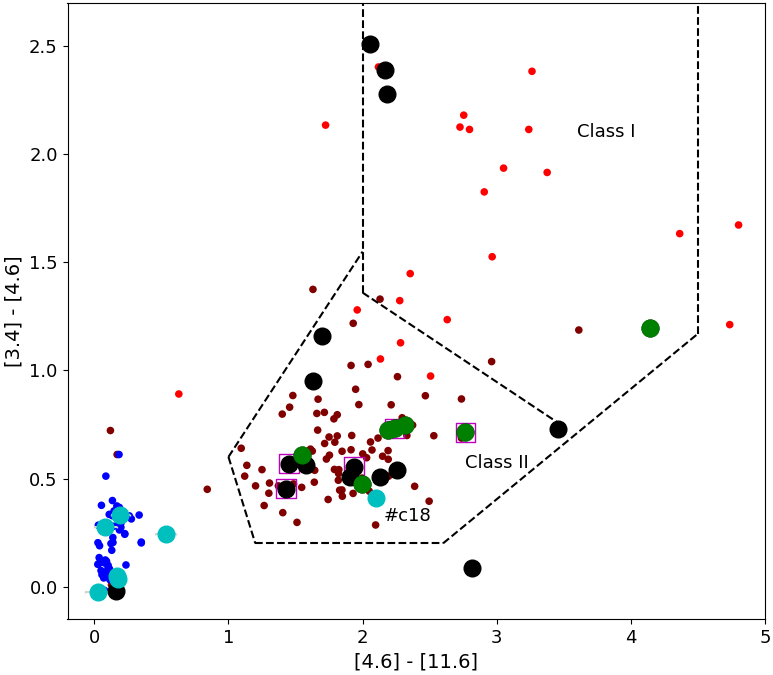}
    \caption{ The {\it WISE} CC diagram for the newly identified comoving sources shown using filled cyan circles. The known YSO candidates detected (green circle) and not detected (black circle) by the \textit{Gaia} are shown using filled circles in green black filled circles. The Class I, II and III sources taken from \citet{2010ApJS..186..259R} are shown in red, maroon and blue dots respectively. The known YSO candidates and the comoving sources spatially found to be associated with the \textit{XMM-Newton} X-ray detection are marked using square boxes in magenta. The 4 sources observed by us spectroscopically are marked. The dashed lines are the criteria used by \citet{2014ApJ...791..131K} to separate the regions occupied by the Class I and Class II sources.}
    \label{fig:wise_CC}
\end{figure}
%**************************************************************************************

The \textit{Spitzer} IRAC (3.6, 4.5, 5.8 and 8.0 $\mu$m) flux values given in \cite{2009ApJS..185..198K} for the known YSO candidates are converted to their respective magnitudes and are shown in the [3.6]-[4.5] vs. [5.8]-[8.0] CC diagram in Fig. \ref{fig:spitzer_CC} using filled circles in green (with reliable \textit{Gaia DR2} data) and black (without reliable \textit{Gaia} DR2 data). Of the 20 YSO candidates with reliable \textit{Gaia} DR2 data, we obtained \textit{Spitzer} IRAC magnitudes for 15 sources. Of the remaining 38 YSO candidates, 28 are associated with reliable \textit{Spitzer} IRAC magnitudes. Since the IRAC magnitudes for the comoving sources are not available in the literature, we obtained their magnitudes by performing photometry using MOPEX. Nine comoving sources are found to lie outside the \textit{Spitzer} FOV. Four comoving sources do not have reliable photometric flux in 8.0 $\mu$m band. None of the comoving sources show emission in MIPS 24 $\mu$m. Therefore we show only IRAC magnitudes in Table \ref{tab:2mass_wise_yso_com}. The remaining seven sources for which we have reliable data are shown in  Fig. \ref{fig:spitzer_CC} using filled circles in cyan. We show the boundaries of Class I, II and III sources adopted from \cite{2009A&A...504..461F,2010ApJ...717.1067C} in Fig. \ref{fig:spitzer_CC}. The three new comoving sources are found to occupy the region generally occupied by the Class III sources.

The \textit{WISE} magnitudes for the sources are also obtained from the \citet{2014yCat.2328....0C} catalogue by making a search around each of them with a search radius of 3$^{\prime\prime}$. The sources having photometric quality `A' (SNR$\geqslant$10) in \textit{W1}, \textit{W2} and \textit{W3} bands are selected as for a majority of the sources, the magnitudes given in the \textit{W4} band are only upper limits. We confirmed the detection of the sources in each bands by making visual inspection of the \textit{WISE} images in all the bands. We noticed that for a notable number of sources, though the catalogue provides magnitude values in 11.6 and 22.1 $\mu$m bands, on visual inspection, no detection was found on the images. The search results are shown in the [3.4]$-$[4.6] vs. [4.6]$-$[11.6] CC diagram (Fig. \ref{fig:wise_CC}). The meaning of the symbols are same as in Fig. \ref{fig:spitzer_CC}. Of the 20 comoving sources, we found a counterpart for 7 in the \textit{WISE} database. Among these, one source is common with those shown in Fig. \ref{fig:spitzer_CC}. Six additional sources, including \#c18, are found in the \textit{WISE} database. Of 20 YSO candidates with reliable \textit{Gaia} DR2 data, we obtained \textit{WISE} magnitudes for 11 sources. Both, the known YSO candidates having reliable \textit{Gaia DR2} data and the comoving sources are presented in Table \ref{tab:2mass_wise_yso_com}. Of the remaining known YSO candidates having no reliable data or the \textit{Gaia} detection, we obtained \textit{WISE} magnitudes for 10 sources which are presented in Table \ref{tab:NRI_other}. In both Fig. \ref{fig:spitzer_CC} and Fig. \ref{fig:wise_CC} we again show Class I, II and III (red, maroon and blue dots, respectively) sources taken from \citet{2010ApJS..186..259R} for comparison. 

The difference in the distribution of the known YSO candidates and the comoving sources is more prominent in Fig. \ref{fig:spitzer_CC} and in Fig. \ref{fig:wise_CC}. It is evident that the known YSO candidates fall in a region generally occupied by Class I and Class II objects. The sources showing spatial association with the X-ray detection are marked with square boxes in magenta. A number of the newly identified comoving sources fall in a region populated by the Class III sources. The comoving source \#c18 is located in a region occupied by the Class II sources in \textit {WISE} CC diagram. This is the source which shows second highest EW(H$\alpha$). The source \#c9 which shows a filled-in emission in H$\alpha$ is located in the Class I/II region and the source \#c14 which shows smallest EW(H$\alpha$), occupies Class III region in \textit{Spitzer} CC diagram. The source \#c8, which shows highest EW(H$\alpha$), is not shown in both Fig. \ref{fig:spitzer_CC} and Fig. \ref{fig:wise_CC} because of its non-detection by the \textit{Spitzer} IRAC and the \textit{WISE}.

Owing to the lack of reliable photometry in all the bands, we were unable to classify a significant number of comoving sources into various classes based on the CC diagrams. Thus we made an attempt to classify them by calculating their spectral index values. The spectral index, $\alpha$, is defined as, $\alpha$ = $\frac{dlog(\lambda F_{\lambda})}{dlog(\lambda)}$; where F$_{\lambda}$	denotes the flux density at wavelength $\lambda$. We calculated the index by a least-squares fit to all the available data from 2MASS K$_{s}$ (2.2 $\mu$m) to \textit{WISE} 22 $\mu$m/MIPS 24 $\mu$m. \cite{1994ApJ...434..614G} proposed the following criteria to classify sources into different evolutionary classes: Class I: $\alpha \geq$ 0.3, Flat Spectrum: 0.3 $> \alpha \geq$ -0.3, Class II: -0.3 $> \alpha \geq$ -1.6, and Class III: $\alpha \leq$ -1.6. A decrease in the value of $\alpha$ implies that the amount of the circumstellar material around the sources decreases as they evolve. The SEDs and the spectral index values of the \textquote{flat} spectrum sources are intermediate between those of deeply embedded \textquote{Class I} and those of more evolved Class II YSOs. We obtained spectral index values of the known YSO candidates from \cite{2009ApJS..185..198K} shown in Table \ref{tab:2mass_wise_yso_com} and \ref{tab:NRI_other}. For five YSO candidates that are not listed in the catalogue by \cite{2009ApJS..185..198K}, we calculated their $\alpha$ values from the available photometric data in various bands. Thus, based on the $\alpha$ values, of the 58 YSO candidates, 7 are classified as Class I, 4 as flat spectrum sources, 35 as Class II and 6 as Class III sources. Of the 20 comoving sources, 4 are classified as Class II and 12 as Class III sources. Remaining 6 YSO candidates and 4 comoving sources are unclassified due to the lack of reliable photomeric data.

%#####################################################################################

\subsubsection{Optical and near-IR colour-magnitude diagrams}

Using $G$ and $G_{\mathrm{RP}}$ magnitudes of 20 YSO candidates and 20 comoving sources from the \textit{Gaia} DR2, we constructed M$_{G}$ vs. ($G$-$G_{\mathrm{RP}}$) colour-magnitude diagram (CMD). We converted the G magnitudes into absolute magnitudes using the expression M$_{G}$ = $G$ + 5 - 5~log($d$), where $d$ is the distance taken from \citet{2018AJ....156...58B}. Out of 20 YSO candidates for which we have reliable distances from the \textit{Gaia} DR2 (Table \ref{tab:YSO_gaia}), extinction values for 11 sources are available in \cite{2009ApJS..185..451K}. Based on our spectral type determination, only 4 of the 20 comoving sources have extinction values. As the individual extinction values for most of the sources are unknown, we constructed M$_{G}$ vs. ($G$-$G_{\mathrm{RP}}$) CMD without correcting for the extinction. We did not use extinction values provided by the \textit{Gaia} DR2 catalogue as these values are not accurate at the individual star level \citep{2018A&A...616A...8A}. The effective temperatures and extinctions listed in \textit{Gaia} DR2 are based on a naked stellar model where the contribution from the dust in the star forming region and/or the protoplanetary disc are ignored. 

\begin{figure}
    \includegraphics[width=8.3cm, height=13cm]{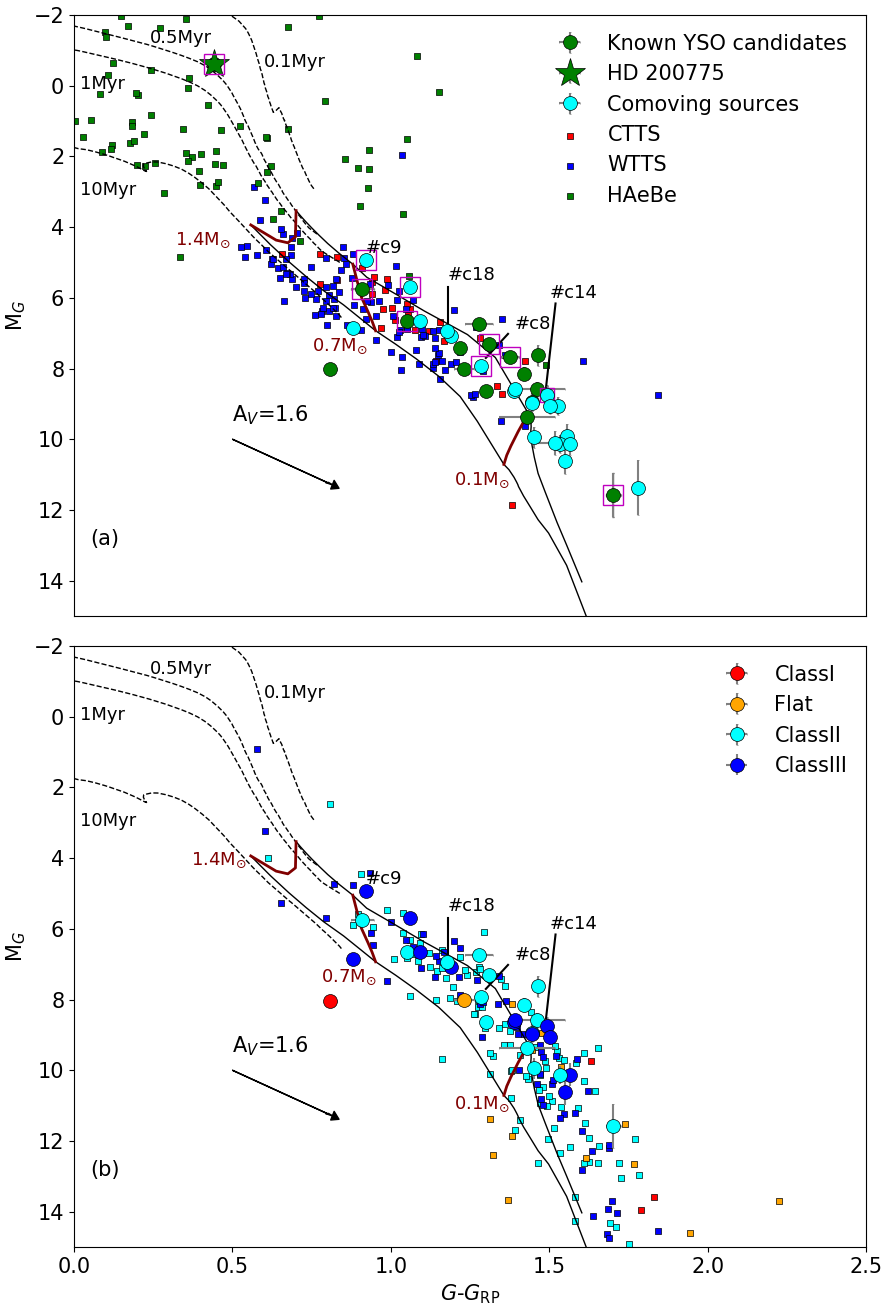}
    \caption{{\bf (a)} The M$_{G}$ vs. ($G$-$G_{\mathrm{RP}}$) colour-magnitude plot of the YSO candidates and the comoving sources in the vicinity of HD 200775. The dashed lines indicate the isochrones from PARSEC models \citep{2017ApJ...835...77M} and solid curves represent the same from CIFIST models \citep{2015A&A...577A..42B}. Green circles: known YSO candidates, cyan circles: newly identified comoving sources, magenta square boxes: X-ray sources. The comoving sources observed spectroscopically are marked. The arrow represents extinction of 1.6 magnitude \citep[the average extinction toward NGC 7023 estimated by][]{2009ApJS..185..451K}. {\bf (b)} Same as above but the sources are shown according to their classifications. Red square boxes: Class I, orange square boxes: Flat spectrum, cyan square boxes: Class II and blue square boxes: Class III sources.}
    \label{fig:cm_figure}
\end{figure}

In Fig. \ref{fig:cm_figure} (a) we show the M$_{G}$ vs. ($G$-$G_{\mathrm{RP}}$) CMD for the known YSO candidates (filled circles in green) and the newly identified comoving sources (filled circles in cyan) found in the vicinity of HD 200775. The known YSO candidates and the comoving sources that are showing X-ray emission are also marked using magenta square boxes. Also shown are HAeBe \citep{1994A&AS..104..315T}, CTTS and WTTS \citep{2010ApJ...724..835W} using dots in green, red and blue, respectively. A reddening vector corresponding to a median extinction of 1.6 magnitude estimated by \cite{2009ApJS..185..451K} for NGC 7023 is shown in Fig. \ref{fig:cm_figure} (a) and (b). The reddening vector is computed based on \cite{2019A&A...623A.108B}. The PMS isochrones corresponding to 0.1 Myr, 0.5 Myr, 1 Myr and 10 Myr are also shown. We used two grids of models, the CIFIST 2011\_2015\footnote{\url{phoenix.ens-lyon.fr/Grids/BT-Settl/ CIFIST2011\_ 2015/ISOCHRONES/}} models for low mass stars \citep[thick curves in black, ][]{2015A&A...577A..42B}, and the PADOVA tracks Parsec 3.3\footnote{\url{stev.oapd.inaf.it/cmd}} for the higher-mass stars \citep[dashed curves in black, ][]{2017ApJ...835...77M}. Also shown are the positions of HAeBe, CTTS and WTTS sources taken from \citet{1994A&AS..104..315T} and \citet{2010AJ....140.1868W}. The four comoving sources for which we have carried out optical spectroscopy are identified and labeled.

%*********************************************************************************
%*********************************************************************************
\begin{figure}
    \includegraphics[width=8.3cm, height=13cm]{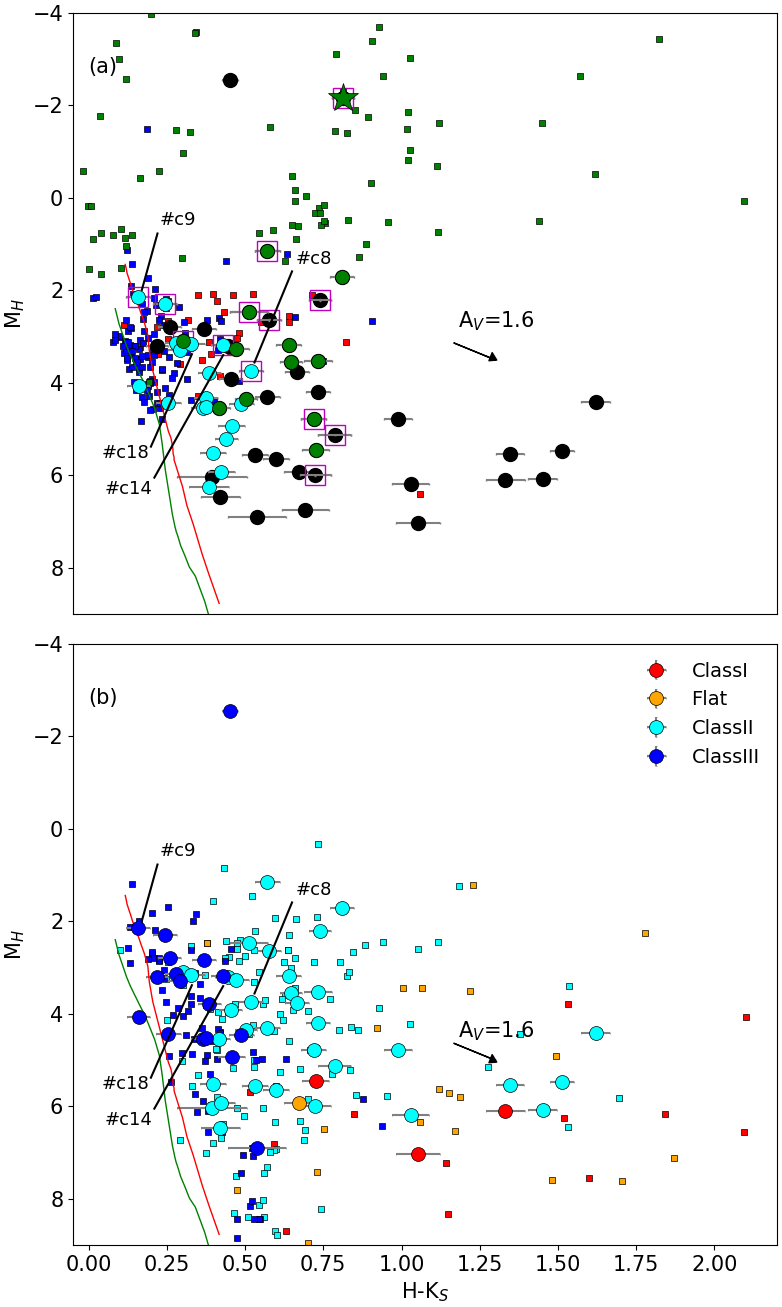}
    \caption{{\bf (a)} M$_{H}$ vs. ($H-K_{S}$) colour-magnitude plot for sources identified in the vicinity of HD 200775. The PMS isochrones corresponding to 1 Myr (red curve) and 10 Myr (green curve) taken from CIFIST models are drawn. Cyan filled circles: newly identified comoving sources, green filled circles: known YSO candidates having reliable \textit{Gaia} DR2 detection, black filled circles: known YSO candidates without reliable or no \textit{Gaia} DR2 detection, green dots: HAeBe, red dots: CTTS, blue dots: WTTS, green star: HD 200775, magenta square boxes: X-ray sources. The comoving sources observed spectroscopically are marked. {\bf (b)} Same as above but sources are shown according to their classifications. Red square boxes: Class I, orange square boxes: Flat spectrum, cyan square boxes: Class II and blue square boxes: Class III sources.}
    \label{fig:cm_figure_2mass}
\end{figure}
%*********************************************************************************

In Fig. \ref{fig:cm_figure} (a), a number of known YSO candidates and comoving sources are lying on or above the 1 Myr isochrone. HD 200775 is found to be of $\sim0.5$ Myr old which is consistent with the previous studies \citep{2008MNRAS.385..391A, 2018A&A...620A.128V, 2019AJ....157..159A}. In Fig. \ref{fig:cm_figure} (b) we show M$_{G}$ vs. ($G$-$G_{\mathrm{RP}}$) CMD of the sources that are classified as Class I, II, III and flat spectrum sources in red, orange, cyan and blue filled circles respectively. Also shown are the locations of Class I, II, III and the flat spectrum sources taken from \cite{2010ApJS..186..259R}, using filled square boxes in red, orange, cyan and blue colors respectively. These are the identified members of the Taurus molecular cloud whose average age is estimated to be $\sim1$ Myr \citep{1992AJ....104..762G}. A total of 187 out of 257 sources have a \textit{Gaia} DR2 counterpart found within a search radius of 3$^{\prime\prime}$. M$_{G}$ values of these sources are corrected for their distances obtained from the \textit{Gaia} DR2. All but two sources identified in the vicinity of HD 200775 show a well defined sequence roughly following the 1 Myr isochrone which is consistent with the median age of the YSO candidates ($\sim$1.6 Myr) obtained by \cite{2009ApJS..185..451K}. The distribution is found to be similar to the sources found in the Taurus molecular cloud. As the extinction vector is parallel to the isochrones for the sources having M$_{G}\lesssim$ 8 magnitude, the extinction will not affect their distribution. However, for sources having M$_{G}\gtrsim$ 8 magnitude, the effect of extinction can cause erroneous age estimates of the sources as can be noticed in Fig. \ref{fig:cm_figure} (b). The source falling below the 10 Myr isochrone is classified by us as a Class I source. This is the only Class I source that shows reliable data in \textit{Gaia} DR2. This source, \#15 (Table \ref{tab:YSO_gaia}), is identified by \citet{2009ApJS..185..451K} as NGC 7023 RS 10 and they classified it as a variable source. They also found this source to be peculiar as it was showing an age older than 10$^{8}$ year. This source is at a distance of 323$^{+4}_{-5}$ pc and the proper motion values (7.174$\pm$0.079 mas/yr in RA and -1.738$\pm$0.075 mas/yr in Dec) are quite consistent with those of other known YSO candidates. As suggested by \citet{2009ApJS..185..451K}, this source could be highly variable. 

In Fig. \ref{fig:cm_figure_2mass} (a) and (b) we show M$_{H}$ vs. ($H-K_{S}$) CMD for the sources seen in the vicinity of HD 200775. Considering that a significant number of known YSO candidates and the comoving sources have 2MASS data, it is possible to compare the properties of the comoving and the known YSO candidates including those having no reliable or no \textit{Gaia} DR2 data. The symbols are identical as in Fig. \ref{fig:cm_figure} (a) and (b) respectively except that the known YSO candidates without reliable or no \textit{Gaia} DR2 data are shown using filled circles in black. The known YSO candidates and the comoving sources found associated with the X-ray sources are also marked using magenta square boxes. The PMS isochrones corresponding to 1 Myr and 10 Myr taken from the CIFIST models are drawn as red and green curves respectively. A reddening vector corresponding to A$_{V}$=1.6 magnitude is drawn. In Fig. \ref{fig:cm_figure_2mass} (a) we also show the HAeBe \citep{1994A&AS..104..315T}, CTTS and WTTS \citep{2010ApJ...724..835W} using squares in green, red and blue, respectively. Evidently, the known YSOs having no reliable or no \textit{Gaia} DR2 data are relatively fainter and show higher values of ($H-K_{S}$) colors. The faintness of these sources could be because of the extinction suffered by them due to the presence of both interstellar and circumstellar material along the line of sight. 

In Fig. \ref{fig:cm_figure_2mass} (b) we show the sources classified based on their spectral index values. As in Fig. \ref{fig:cm_figure} (b), here also we show the Class I, II, III and Flat spectrum sources obtained from \cite{2010ApJS..186..259R}. These sources, associated with the Taurus molecular clouds, show a sequence with the Class I sources lying relatively far away from the 1 Myr isochrone and the Class III sources lying closer to it. The flat spectrum and the Class II sources are found to occupy a region between Class I and Class III sources. Except for the 4 sources that are classified as Class II but showing relatively large ($H-K_{S}$) colours, other sources classified as Class I, II, III and Flat spectrum sources in the vicinity of HD 200775 are also found to follow the sequence shown by the YSOs in the Taurus molecular cloud. Of the 55 sources for which we made the classification, we found 3 Class I (5\%), 34 Class II (62\%), 1 flat spectrum (2\%) and 17 Class III (31\%) sources towards HD 200775. Based on the classification made by \cite{2010ApJS..186..259R} for the sources associated with the Taurus molecular cloud, there are 14 Class I (6\%), 122 Class II (53\%), 20 flat spectrum (8\%) and 76 Class III (33\%) sources. These are the 232 sources out of 257, for which we obtained 2MASS counterparts within 3" search radius (following the strategy used by \cite{2010ApJS..186..259R} to get the 2MASS counterparts) and photometric quality of `A' in all the bands. Based on the statistics of the sources found in different categories, it is apparent that the nature of the sources found in the vicinity of HD 200775 and those in the Taurus molecular cloud are very much similar.
%**********************************************************************************

\subsubsection{Spatial distribution of the sources with respect to HD 200775}
%*********************************************************************************

\begin{figure}
    \includegraphics[height=8cm, width=8.3cm]{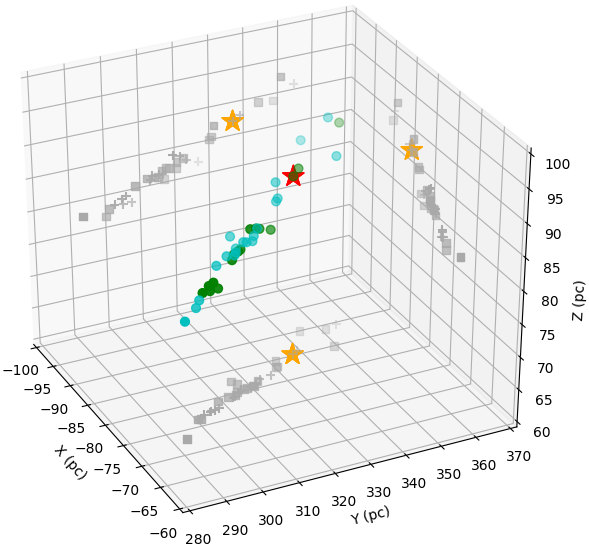}\\
    \caption{The 3D distribution of the known YSO candidates (filled circles in green) and the comoving (filled circles in cyan) sources having \textit{Gaia} DR2 distances. Their respective projections on the X, Y and Z planes are shown using plus and square symbols in black (X positive towards the Galactic center, Y positive towards the Galactic east and Z positive towards the Galactic north. HD 200775 is identified using a star symbol in red. Its projection in all the three planes is shown using a star in orange.}
    \label{fig:3d_figure}
\end{figure}
\begin{figure}
	\includegraphics[height=8.5cm, width=8.3cm]{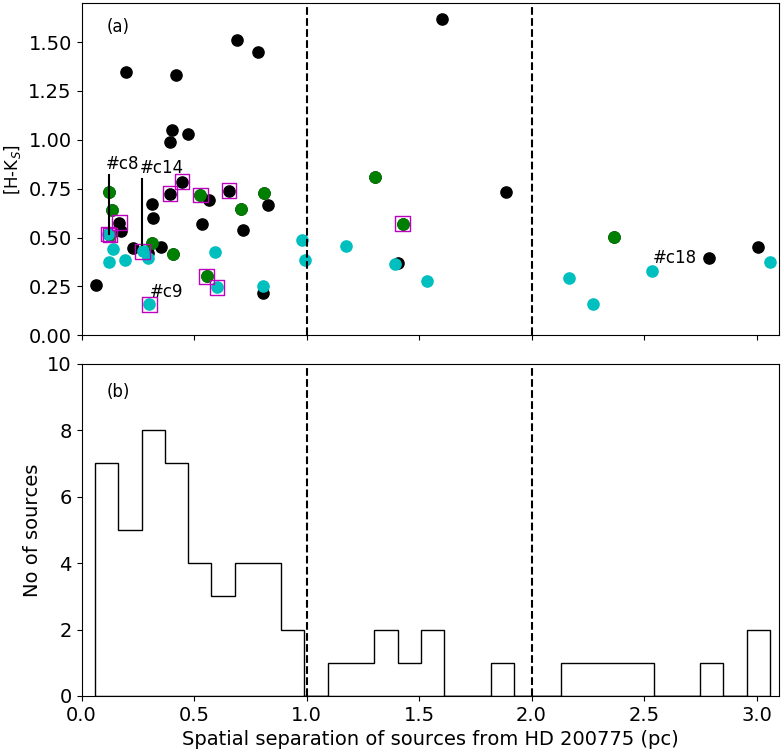}\\
	\caption{(\textbf{a}) The 2D spatial distribution of the known YSO candidates (filled circles in green: sources having reliable \textit{Gaia} data, filled circles in black: sources without \textit{Gaia} data) and the newly identified comoving sources (filled circles in cyan) with respect to the star HD 200775. Their (\textit{H-K$_{S}$}) values are shown in the vertical axis. Sources having X-ray emission are marked with magenta square boxes. The sources observed by us using HFOSC are also marked. (\textbf{b}) Histogram of the distribution of the known YSO candidates and the newly identified comoving sources with respect to the star HD 200775.}
	\label{fig:ang_dist_figure}
\end{figure}

%*********************************************************************************
Now that the distances to a number of sources towards the direction of HD 200775 are known, we constructed a 3-D distribution of the known YSO candidates and the comoving sources in cartesian coordinate system as shown in Fig. \ref{fig:3d_figure}. The known YSO candidates and the comoving sources are shown using filled circles in green and cyan, respectively. The projection of the distribution on the X-Y, Y-Z and X-Z planes are also shown (plus sign for the comoving and square boxes for the known YSO candidates). The X-, Y- and Z-positives are towards the Galactic center, the Galactic east and the Galactic north respectively. The location of HD 200775 is shown using a star symbol. We find a significant ($\sim2\sigma$) shift in the distance of HD 200775 with respect to the distances of other known YSO candidates. Assuming a distance of 335 pc to all the sources (both with and without \textit{Gaia} DR2 data) found in the vicinity of HD 200775, we calculated the projected separation (in parsec) of them with reference to the position of HD 200775. The 2D distribution of 58 sources (both known YSO candidates and comoving sources having good quality data from 2MASS) thus obtained is shown in Fig. \ref{fig:ang_dist_figure}. The distribution shows that the maximum projected extent of the sources on the sky plane is $\sim6$ pc in diameter. If we assume a similar depth along the line of sight direction also, then HD 200775 will lie outside the cloud. This is inconsistent with the fact that the nebulosity, NGC 7023, is illuminated by HD 200775. 

The \textit{Gaia} DR2 distance of 357$^{+6}_{-7}$ pc to HD 200775, though is found to be consistent with the 430$^{+160}_{-90}$ pc estimated using the Hipparcos parallax measurements \citep{1998A&A...330..145V}, the value is found to be significantly different from the recomputed distance of 520$^{+180}_{-110}$ pc made by \cite{2007A&A...474..653V}. HD 200775 is a triple system composed of a double-lined spectroscopic binary at $\sim18$ mas separation \citep{2001ApJ...546..358M} and a third companion at 6$\arcsec$ separation \citep{1994ApJ...433..199L}. Based on the radial velocity measurements and the astrometric data, \citet{2013A&A...555A.113B} estimated a dynamical distance of 320$\pm51$ pc. We note that binarity was not worked into the astrometric solution for the \textit{Gaia} DR2. The astrometric excess noise \citep[$\epsilon_{i}$;][]{2012A&A...538A..78L} which measures the disagreement between the observations of a source and the best-fitting standard astrometric model is found to be 0.185 milli-arcsec for HD 200775. A positive value of $\epsilon_{i}$ signifies that the residuals are statistically larger than expected. The significance of $\epsilon_{i}$ depends on the value of a dimensionless parameter, D \citep[``astrometric\_excess\_noise\_sig";][]{2012A&A...538A..78L}\footnote{\url{https://gea.esac.esa.int/archive/documentation/GDR2/Gaia_archive/chap_datamodel/}}, which if is greater than 2, implies that $\epsilon_{i}$ is significant. The value of D for HD 200775 is 12.3 suggesting that there is a disagreement between the observations and the standard astrometric model. In addition, the RUWE value of HD 200775 is 1.6, indicating that the astrometric solution is not well-behaved. Presence of unresolved binaries is believed to be one of the reasons for such deviations \citep{2018A&A...616A...1G}. Therefore the present \textit{Gaia} DR2 parallax measurement of HD 200775 is possibly inaccurate.

%*********************************************************************************

%*********************************************************************************

In Fig. \ref{fig:ang_dist_figure} (a), along the vertical axis we show the (\textit{H-K$_{S}$}) colour of the sources considering it to be a proxy for the amount of circumstellar material present around each source. The objects identified as X-ray sources are marked with square boxes in magenta. Evidently, the comoving sources show the lowest of the (\textit{H-K$_{S}$}) colour and are spatially well correlated with the known YSO candidates. A majority of the sources (44 or 75\%) are  distributed within $\sim1$ pc distance from HD 200775, 8 ($\sim13$\%) are distributed between 1 and 2 pc and only 7 ($\sim12$\%) are located beyond 2 pc. Both the known YSO candidates and the comoving sources are among the sources distributed within 1 pc from HD 200775. The distribution of the sources presented in Fig. \ref{fig:ang_dist_figure} shows an enhancement in the vicinity of HD 200775 but, instead of a high density clustering, we detect a loose association. Dynamical dissipation was suggested as one of the possible reasons for the lack of clustering in HD 200775 \citep{2001A&A...366..873F}. The distribution of the proper motions of the young stars associated with a region can yield the velocity dispersion provided the distance is well known. Using the distance of 335 pc and the dispersion in the proper motion values in RA and Dec for the known YSO candidates and the comoving sources combined, we calculated a velocity dispersion of $\sim1$ km s$^{-1}$ for the sources found in the vicinity of HD 200775. Given the velocity dispersion and an age of $\sim1$ Myr, the sources around HD 200775 could have moved by $\sim$1 pc from where they were born. If we assume that they were all born very close to HD 200775, the distribution seen in Fig. \ref{fig:ang_dist_figure} is quite consistent. But in such a case, we expect the distribution of the known YSO candidates and the comoving sources around HD 200775 to be symmetrical.

%*********************************************************************************

\begin{figure*}
    \includegraphics[height=8.5cm, width=\textwidth]{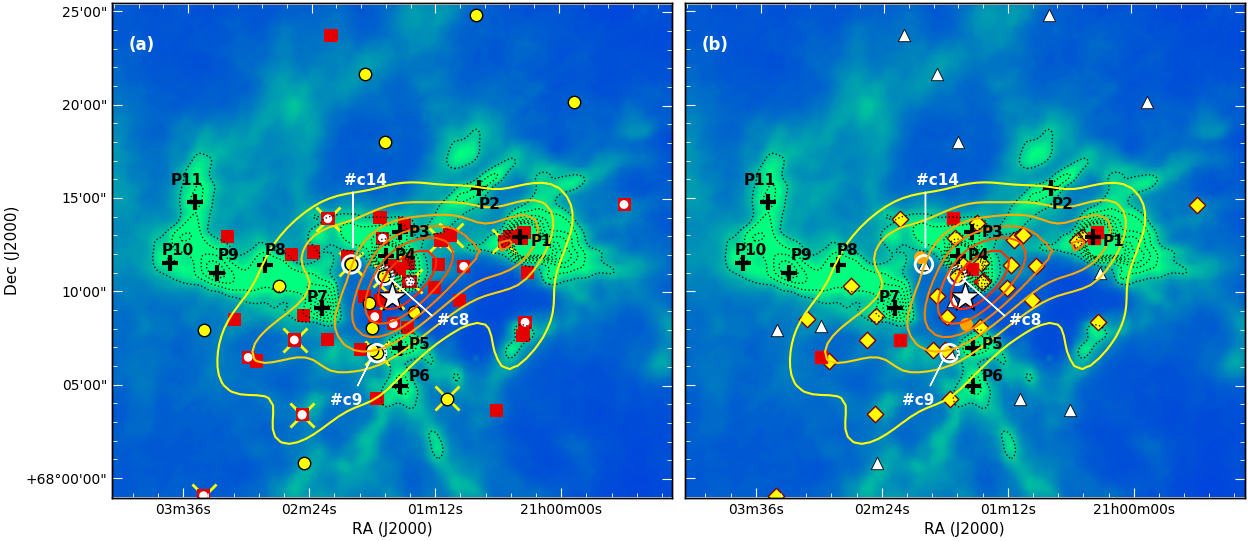}
    \caption{{\bf (a)} The spatial distribution of the YSO candidates and the comoving sources overplotted on the $Herschel$ column density map of the region surrounding HD 200775. The position of HD 200775 is identified using a white star symbol. Peak emission regions are shown in black `+' symbols. Red square: known YSO candidates; Red square + white dot: known YSO candidates with reliable \textit{Gaia} DR2 data; Yellow filled circle: comoving sources identified in this work; Yellow filled circle + open circle in white: comoving sources observed by us spectroscopically; Yellow cross: X-ray detection by the \textit{XMM-Newton} satellite. Black Plus: Column density peaks. The contours shown in thick lines represent the source distribution around HD 200775. {\bf(b}) Represents the same as (a) but with the classifications of the sources. Red square box: Class I, Orange filled circle: Flat spectrum, Yellow diamond: Class II, White filled triangle: Class III sources.}\label{fig:c18o_YSO}
\end{figure*}
%*********************************************************************************

In Fig. \ref{fig:c18o_YSO} (a) we show the distribution of 59 sources (YSO candidates + comoving sources) that fall within the \textit{Herschel} dust column density map. The known YSO candidates are identified using red square symboles. Of these, the sources having no reliable \textit{Gaia} DR2 data are identified with red square + white dot. The comoving sources are shown using circles in yellow. Both the known YSO candidates and the comoving sources associated with the X-ray detection are identified using cross symbols. The contours shown in thick lines represent the source distribution around HD 200775 (white star symbol). The distribution of the sources is not symmetrical with respect to HD 200775. At immediate surroundings of HD 200775, source distribution is concentrated more towards the north and as we move further out, sources are sparsely distributed more in the east-west direction. \citet{1999A&A...342..515T} found a correlation between the spectral type of HAeBe stars and the stellar density surrounding them. However, HD 200775 was shown as an exception due to the lack of clustering around it. \citet{2001ASPC..243..377T} discussed the results in connection with two competing models, namely, physical \citep{1998MNRAS.298...93B} and random sampling models \citep{1999ApJ...515..323E} which are advocated to explain the formation of high-mass stars. According to the physical models, the presence of clustering is a pre-requisite for the formation of massive stars but on the other hand, they are unrelated according to the random models. In fact, the presence of a few of the massive stars (like HD 200775) not surrounded by an enhanced stellar density was shown as a support to the random models. Even with the identification of 20 additional comoving sources in the vicinity of HD 200775, because these sources are more sparsely distributed compared to the known YSO candidates, the stellar density around HD 200775 is still very low and hence favours the random sampling models.

HD 200775 is residing inside a biconical cavity, which is believed to have been carved out in an earlier evolutionary stage by a currently inactive bipolar outflow \citep{1996A&A...310..286F}. The outflow axis runs along the east-west direction. \citet{1998A&A...334..253F}, based on the morphology of the parent cloud, classified HAeBe stars into three groups. They classified HD 200775 as Type III which implies that the source has completely dispersed the surrounding dense gas and is currently located inside a cavity of molecular cloud. The expulsion of gas from the vicinity of OB stars could limits the amount of molecular material available for the formation of additional stars which could be a possible reason for the lack of clustering around HD 200775. Using the CLUMPFIND \citep{1994ApJ...428..693W} task on the dust column density map made using the \textit{Herschel} images, we found a total of 11 peak emission regions (P1 - P11) which are shown in Fig. \ref{fig:c18o_YSO}. The contours are plotted from $3.2\times10^{21}$ to $1.1\times10^{22}$ cm$^{-2}$ in steps of $1.3\times10^{21}$ cm$^{-2}$. While two peak emission regions are located to the north and two to the south of HD 200775, the rest of the peaks are distributed mostly along the east-west direction. This roughly coincides with the orientation of the biconical cavity \citep{1996A&A...310..286F}. The velocity-intensity map of $^{13}$CO~J=2-1 gas emission \citep{2013MNRAS.429..954Y} and the dust emission (Fig. \ref{fig:c18o_YSO}) surrounding HD 200775 show a large intensity gradient along the sides facing HD 200775. It is possible that the bipolar outflow and the stellar wind from HD 200775 might have been responsible for shaping the geometry of the material seen around it.

A significant number of the sources are found to be associated with the peaks towards P1, P3 and P4. Evidence of current star formation is seen towards P1 \citep[e.g., ][]{2013AJ....145...35R}. Using $^{12}$CO J=3-2 observations, \cite{2013MNRAS.429..954Y} detected a non-Gaussian line profile with the peak skewed to the blue indicative of infall motions. Presence of infall motions suggests that star formation is still active. In Fig. \ref{fig:c18o_YSO} (b), we show the distribution of sources classified as Class I, II, III and flat spectrum sources. Of the six sources classified as Class I, five (squares in red) are located within the mapped area. Among these, three sources are located towards P1 and P4. Possibly, star formation in L1172/1174 complex might have begun roughly $\sim1$ Myr ago but the very young sources, particularly those of Class I type, which are associated with the P1 and P4 peaks might have been formed as a result of positive feedback from HD 200775. P1 is found to be associated with two Class I sources and a number of outflows \citep{2009ApJS..185..198K, 2013AJ....145...35R}. A dynamical age of $\sim$7500 yr is estimated by \cite{2013AJ....145...35R}. This suggests that some of these young sources might have been formed very recently. The presence of infall motions in Core 1 (P1 in Fig. \ref{fig:c18o_YSO}) and intensity gradient to the south of Core 3 (P3 and P4 in Fig. \ref{fig:c18o_YSO}) shown by \cite{2013MNRAS.429..954Y} supports this argument. Based on this and the shock features observed at 4.5$\mu$m, they suggested that star formation around HD 200775 could have been possibly triggered due to HD 200775. Star formation scenario in the Cepheus region was presented by \cite{2009ApJS..185..451K}. However, to understand the history of star formation, it is important to study the true nature of the rest of the comoving sources and study the dynamical state of the gas surrounding HD200775.

%**********************************************************************************

%####################################################################################

\section{Conclusions}\label{sec:con}

The region surrounding the well-known reflection nebula, NGC 7023, illuminated by a Herbig Be star, HD 200775, located in the dark cloud L1174 is known for low to intermediate-mass star formation activity. Distance to L1172/1174 complex of which L1174 is a part is not very well constrained. In this work, using \textit{Gaia} DR2 distances and proper motion values, we determined distance to the cloud using the known YSO candidates. We also used polarization measurements of 545 stars that are projected on the cloud to estimate the same. Then we searched for additional comoving sources in the vicinity of HD 200775. The main results from this study are summarized below: 

\begin{itemize}
    \item Based on the distances and proper motion values of 20 previously known YSO candidates associated with L1174 we obtained a distance of $335\pm11$ pc to the cloud. The distance estimated for the entire cloud using the polarization measurements of the foreground and background stars was also found to be consistent with the $\sim335$ pc distance. 
    \item With the knowledge of distance and proper motion values of the YSO candidates associated with the cloud, we searched for additional comoving sources in the vicinity (1$\degr\times1\degr$ area) of HD 200775. We found a total of 20 sources which are comoving with the known YSO candidates. Compared to the known YSO candidates, the comoving sources are found to show less IR excess. From their locations on the CMDs (optical and near-IR) and by comparison with the PMS isochrone models, they are found to be young ($\sim$ 1 Myr) sources. 
    \item Based on the \textit{XMM-Newton} data, we found 19 reliable X-ray detection around HD200775. Of these, 10 are found to be associated with the known YSO candidates and 4 with the newly identified comoving sources.
    \item We made spectroscopic observations of 4 of the 20 comoving sources of which three are X-ray sources. We found H$\alpha$ in emission in 3 of the 4 sources.
    \item We classified the YSO candidates and the comoving sources based on their spectral index values as Class I, II, III and flat spectrum sources. A majority of the comoving sources are classified as Class III. About 80\% of the known YSO candidates and the comoving sources are found to be distributed within $\sim1$ pc distance from HD 200775. 
    \item Spatial correlation of a number of Class I sources with the density peaks of dust identified from the \textit{Herschel} map suggests that star formation is currently active in the vicinity of HD 200775 and some of these young sources could have been formed as a result of positive feedback from it.
\end{itemize}
A detailed study of all the identified sources, especially, the 20 newly identified comoving sources is required to estimate their spectral type, extinction and luminosity to confirm their true characteristics. 

\section*{Acknowledgements}

We thank all the supporting staff at ARIES, Nainital and the staff of IAO, Hanle and CREST, Hosakote, who made these observations possible. The facilities at IAO and CREST are operated by the Indian Institute of Astrophysics, Bangalore. C.W.L was supported by the Basic Science Research Program (2019R1A2C1010851) through the National Research Foundation of Korea. This work has made use of data from the following sources: (1) European Space Agency (ESA) mission {\it Gaia} (\url{https://www.cosmos.esa.int/gaia}), processed by the {\it Gaia} Data Processing and Analysis Consortium (DPAC, \url{https://www.cosmos.esa.int/web/gaia/dpac/consortium}). Funding for the DPAC has been provided by national institutions, in particular the institutions participating in the {\it Gaia} Multilateral Agreement; (2) the Two Micron All Sky Survey, which is a joint project of the University of Massachusetts and the Infrared Processing and Analysis Center/California Institute of Technology, funded by the National Aeronautics and Space Administration and the National Science Foundation; (3) the {\it Herschel} SPIRE images from {\it Herschel} Science Archive (HSA). \textit{Herschel} is an ESA space observatory with science instruments provided by European-led Principal Investigator consortia and with important participation from NASA; (4) the NASA/ IPAC Infrared Science Archive, which is operated by the Jet Propulsion Laboratory, California Institute of Technology, under contract with the National Aeronautics and Space Administration; (5) the Wide-field Infrared Survey Explorer, which is a joint project of the University of California, Los Angeles, and the Jet Propulsion Laboratory/California Institute of Technology, funded by the National Aeronautics and Space Administration; (6) the $XMM-Newton$, an ESA science mission with instruments and contributions directly funded by ESA Member States and NASA; and (7) the SIMBAD database is operated at CDS, Strasbourg, France.

%###################################################################################################
\bibliographystyle{mnras}
\bibliography{reference} 
%###################################################################################################
\begin{landscape}
    \begin{table}
    \caption{Distance and proper motion values of the known YSO candidates identified towards L1172/1174 from \textit{Gaia} DR2.}
    \label{tab:YSO_gaia}
    \renewcommand{\arraystretch}{1.3}
    \begin{tabular}{lllcllllccccccc} % four columns, alignment for each 
    \hline
    \#& RA  & Dec  & Source & Distance & $\mu_{\alpha\star}$ ($\Delta\mu_{\alpha\star}$) & $\mu_{\delta}$ ($\Delta\mu_{\delta}$) & G (eG) & RUWE & Spectral & H$\alpha$& 2MASS & $Spitzer$ & $WISE$ & $X-ray$ \\
     & ($^{\circ}$) & ($^{\circ}$) & ID & (pc)& (mas/yr) & (mas/yr) & (mag) & & Type & EW(\AA) & & & &\\
     (1)&(2)&(3)&(4)&(5)&(6)&(7)&(8)&(9)&(10)&(11)&(12)&(13)&(14)&(15)\\
    \hline 
    1   & 314.845315    & 68.245467 & 2270256357606156160 & 339$^{+5}_{-5}$ & 7.333$\pm$0.087 & -1.599$\pm$0.079 & 14.3789$\pm$0.0150 & 1.1 & K1 & -9.6 & \checkmark & \checkmark &\checkmark &-\\ 
    2   & 315.084447    & 68.140772 & 2270240208529136128 & 341$^{+10}_{-10}$ & 7.630$\pm$0.147 & -1.182$\pm$0.158 & 17.0301$\pm$0.0309 &1.1 &- &- & \checkmark & \checkmark &\checkmark &-\\
    3   & 315.231481    & 68.190885 & 2270240723925216640 & 333$^{+12}_{-13}$ & 7.205$\pm$0.184 & -1.649$\pm$0.280 & 16.5545$\pm$0.0020 & 1.3 &-&- & \checkmark & \checkmark &\checkmark &-\\ 
    4    & 315.359984   & 68.177338 & 2270245637367812352 & 325$^{+7}_{-6}$ & 7.652$\pm$0.104 & -1.416$\pm$0.121 & 14.8691$\pm$0.0049 & 1.2 & M0IV & -59.8 & \checkmark & \checkmark &\checkmark & \checkmark\\
    5   & 315.362884    & 68.177214 & 2270245637367812608 & 360$^{+13}_{-13}$ & 6.667$\pm$0.174 & -0.865$\pm$0.189 & 16.3712$\pm$0.0266 & 1.5 & M0 & -63.9 & \checkmark & - & - & -\\ 
    6$^{*}$ & 315.399160 & 68.223752 & 2270246874315480960 & 277$^{+27}_{-34}$ & 11.343$\pm$0.673 & -3.039$\pm$0.706 & 17.7558$\pm$0.0107 & 3.3&- & -& \checkmark & \checkmark &\checkmark &-\\ 
    7   & 315.400352    & 68.139576 & 2270245259410695808 & 332$^{+6}_{-6}$ & 6.971$\pm$0.106 & -1.638$\pm$0.117 & 15.6235$\pm$0.0130 & 1.4&M0 & -106.0 & - & \checkmark &\checkmark &-\\
    8    & 315.403923   & 68.163263 & 2270245431209611776 & 357$^{+6}_{-5}$ & 8.336$\pm$0.079 & -1.566$\pm$0.083 & 7.1592$\pm$0.0010 & 1.6 &- &- & \checkmark & -&-& \checkmark\\ 
    9    & 315.427117   & 68.215960 & 2270246148467658624 & 323$^{+8}_{-9}$ & 7.530$\pm$0.148 & -0.668$\pm$0.164 & 15.6942$\pm$0.0033 & 1.5&M2 & -28.8  & \checkmark & \checkmark &\checkmark &-\\ 
    10   & 315.444875   & 68.145894 & 2270057689598399360 & 335$^{+5}_{-4}$ & 7.770$\pm$0.071 & -1.428$\pm$0.078 & 15.0363$\pm$0.0057 & 1.1&M1 & -64.5 & \checkmark & \checkmark &- &-\\
    11   & 315.558650   & 68.233141 & 2270247492790812928 & 321$^{+26}_{-31}$ & 8.666$\pm$0.528 & -1.736$\pm$0.568 & 19.1129$\pm$0.0059 & 1.1&-&- & \checkmark & \checkmark & \checkmark & \checkmark\\ 
    12$^{*}$ & 315.584979 & 68.423341 & 2270276041439733504 & 277$^{+19}_{-22}$ & 7.262$\pm$0.483 & -3.589$\pm$0.530 & 14.6337$\pm$0.0158 &7.9&- & -& \checkmark & - & \checkmark &-\\
    13   & 315.617758   & 68.058287 & 2270056246489398528 & 343$^{+6}_{-6}$ & 6.868$\pm$0.098 & -1.319$\pm$0.107 & 15.3618$\pm$0.0053 & 1.1&K7 & -56.0 &- & \checkmark & \checkmark & \checkmark\\
    13b  & 315.618244    & 68.057669 & 2270056246489398400 & 376$^{+16}_{-15}$ & 8.041$\pm$0.193 & -1.202$\pm$0.215 & 15.5062$\pm$0.0010 & 1.5&M4.5IV & -7.5 &- & -& - & -\\
    14  & 315.637634    & 68.124746 & 2270056968043900928 & 336$^{+3}_{-2}$ & 7.054$\pm$0.044 & -1.017$\pm$0.043 & 14.2794$\pm$0.0078 & 1.0&K5 & -4.0 & \checkmark & \checkmark & \checkmark & \checkmark \\ 
    15  & 315.747855    & 68.108939 & 2270053875667453056 & 323$^{+4}_{-5}$ & 7.174$\pm$0.079 & -1.738$\pm$0.075 & 15.5766$\pm$0.0061 & 1.5&K4 & -8.1 & \checkmark & \checkmark & \checkmark &-\\ 
    16$^{*}$ & 315.748555 & 68.136623 & 2270058303777440640 & 401$^{+85}_{-149}$ & 5.678$\pm$0.849 & -11.026$\pm$1.196 & 16.5831$\pm$0.0108 & 4.7&- & -&\checkmark & \checkmark & \checkmark &-\\
    17  & 315.851719    & 67.985134 & 2270052256462948736 & 326$^{+8}_{-8}$ & 7.269$\pm$0.140 & -2.330$\pm$0.128 & 13.3182$\pm$0.0111 & 5.9 &K2 & -13.1 & \checkmark & \checkmark & \checkmark & \checkmark\\ 
    18$^{*}$ & 315.923249 & 68.396018 & 2270272330587968256 & 559$^{+128}_{-230}$ & 5.407$\pm$0.780 & -3.269$\pm$0.752 & 17.4905$\pm$0.0045 & 4.1&- & -&\checkmark & -& \checkmark &- \\ 
    19  & 315.997585    & 67.824847 & 2270035287049455616 & 346$^{+8}_{-9}$ & 7.722$\pm$0.152 & -2.061$\pm$0.133 & 16.3326$\pm$0.0041 & 1.4&- &- & \checkmark & \checkmark & \checkmark &- \\
    \hline
    \end{tabular}\\
    \renewcommand{\arraystretch}{1}
    Columns 2 \& 3: 2015.5 epoch Right Ascension \& Declination of sources given by \textit{Gaia} DR2.\\
    Column 4: Distance taken from the \citet{2018AJ....156...58B} catalogue.\\
    Column 9: Renormalised Unit Weight Error (RUWE) of the YSO candidates.\\
    Columns 10 \& 11: Spectral type and equivalent width of H$\alpha$ taken from \citet{2009ApJS..185..451K}.\\
    Column 13: X-ray detection by \textit{XMM-Newton}.\\
    Stars \#4 \& \#5 appear as double sources. Resolved by \textit{Gaia} DR2, 2MASS and \textit{Spitzer}. In \textit{WISE} and \textit{XMM-Newton}, they are detected as single source. \textit{WISE} and X-ray are assigned to the brightest source in $K_{s}$ band.\\
    Stars \#13 \& \#13b appear as double sources. Resolved only by \textit{Gaia} DR2. 2MASS classified it as an extended source. Assigned X-ray detection to the brightest source in \textit{Gaia}.\\
    $^{*}$ Sources considered as outliers in our analysis.\\
    
\end{table}
\end{landscape}
\begin{landscape}
    \begin{table}
    \hspace*{-2cm}
    \caption{Distance and proper motion values of the comoving sources from \textit{Gaia} DR2.}
    \label{tab:YSO_new} 
    \renewcommand{\arraystretch}{1.3}
    \scriptsize
    \begin{tabular}{lllcllllcccccccc} % four columns, alignment for each 
    \hline
    $\#$& RA(2015.5) & Dec(2015.5) & Source & Distance & $\mu_{\alpha\star}$ ($\Delta\mu_{\alpha\star}$) & $\mu_{\delta}$ ($\Delta\mu_{\delta}$) & G (eG) & RUWE & Spectral & H$\alpha$&2MASS & $Spitzer$ & $WISE$ & X-ray & Spectral Index\\
     & ($^{\circ}$) & ($^{\circ}$) & ID & (pc) & (mas/yr) & (mas/yr) & (mag) & & Type  &EW (\AA)  & & & & &$\alpha$\\
     (1)&(2)&(3)&(4)&(5)&(6)&(7)&(8)&(9)&(10)&(11)&(12)&(13)&(14)&(15)&(16)\\
    \hline
    %c1  & 314.358914 & 68.184721 & 311$^{+3}_{-2}$      & 6.372$\pm$0.066   & -3.263$\pm$0.041 & 14.321$\pm$0.000 &-&-& \checkmark & \checkmark &-\\
    c1	& 314.358914 & 68.184721 & 2270164445306098560 & 311$^{+3}_{-2}$		& 6.372$\pm$0.066 & -3.263$\pm$0.041 & 14.321$\pm$0.000 & 1.0&-&-& \checkmark & - & \checkmark & - & -2.78\\ 
    c2	& 314.877400 & 67.678993 & 	2269947738436093696 & 369$^{+13}_{-13}$	& 7.472$\pm$0.183 & -3.129$\pm$0.216 & 16.881$\pm$0.002 & 1.0 & -&-& \checkmark & - & \checkmark & - & -2.41\\
    c3  & 314.963955 & 68.336928 & 2270263092114875520 & 373$^{+13}_{-13}$    & 7.315$\pm$0.177  & -1.844$\pm$0.166 & 16.825$\pm$0.002 & 1.0 & -&-& \checkmark & -&\checkmark &-&-2.56\\
    c4  & 315.201051 & 68.415148 & 2270265016260228608	 & 352$^{+5}_{-5}$      & 7.073$\pm$0.074   & -0.752$\pm$0.079 & 14.800$\pm$0.001 & 1.1 & -&-& \checkmark &- &\checkmark &-&-2.66\\
    c5  & 315.271845 & 68.072754 & 2270238795483033728 & 340$^{+3}_{-3}$      & 7.294$\pm$0.048   & -1.576$\pm$0.056 & 13.363$\pm$0.006 & 1.2 & -&-& \checkmark & -&\checkmark & \checkmark & -2.52\\
    c6  & 315.350786 & 68.150243 & 2270245289474206848 & 351$^{+17}_{-16}$    & 8.195$\pm$0.233   & -0.846$\pm$0.271 & 17.622$\pm$0.003 & 1.3 &-&-& \checkmark & -&-&-&-\\
    c7  & 315.420625 & 68.301379 & 2270248798463735808 & 331$^{+9}_{-8}$ &    7.510$\pm$0.142     & -1.452$\pm$0.170 & 16.243$\pm$0.001 & 1.2 &-&-& \checkmark & \checkmark &\checkmark &-&-2.32\\
    c8  & 315.422157 & 68.182326 & 2270245465569124608 & 337$^{+6}_{-6}$ &    7.222$\pm$0.096     & -1.154$\pm$0.105 & 15.564$\pm$0.002 & 1.3 &M1&-15.03& \checkmark & -&\checkmark & \checkmark &-0.73\\
    c9  & 315.439309 & 68.113265 & 	2270057483439969408 & 342$^{+2}_{-3}$ &    7.595$\pm$0.042     & -3.004$\pm$0.046 & 12.600$\pm$0.002 &1.0&K2&-0.69& \checkmark & \checkmark &\checkmark & \checkmark &-2.45\\
    c10  & 315.450292 & 68.115938 & 2270057517799708032 & 338$^{+13}_{-13}$    & 8.084$\pm$0.206   & -1.537$\pm$0.280 & 17.776$\pm$0.002 &1.0&-&-& \checkmark & \checkmark &\checkmark &-&-0.61\\
    c11 & 315.452017 & 68.135740 & 	2270057685302330240 & 350$^{+43}_{-34}$    & 8.426$\pm$0.623   & -2.081$\pm$0.684 & 19.088$\pm$0.004 & 1.3&-&-& \checkmark & -&-&-&-\\
    c12 & 315.458232 & 68.158415 & 2270057723958137600 & 333$^{+13}_{-12}$    & 8.312$\pm$0.202   & -1.153$\pm$0.219 & 16.674$\pm$0.001 & 1.5&-&-& \checkmark & -&-&-&-\\
    c13 & 315.470563 & 68.362746 & 2270252367579936768 & 328$^{+18}_{-17}$    & 8.544$\pm$0.324   & -1.593$\pm$0.369 & 18.204$\pm$0.002 &1.1&-&-& \checkmark & \checkmark &\checkmark &-&-2.30\\
    c14 & 315.501565 & 68.192440 & 2270245809166510080 & 319$^{+11}_{-10}$    & 7.975$\pm$0.177   & -1.400$\pm$0.193 & 16.260$\pm$0.001 &1.3&M3&-4.06& \checkmark & \checkmark &\checkmark & \checkmark & -1.98\\
    c15 & 315.612647 & 68.015216 & 2270053081097108480 & 339$^{+17}_{-14}$    & 7.643$\pm$0.282   & -1.804$\pm$0.294 & 17.782$\pm$0.002 &1.0&-&-& \checkmark & -&\checkmark &-&-2.21\\
    c16 & 315.675742 & 68.172632 & 2270058651671079808 & 334$^{+15}_{-13}$    & 7.885$\pm$0.231   & -1.042$\pm$0.250 & 17.566$\pm$0.005 &1.3&-&-& \checkmark & \checkmark &\checkmark &-&-1.45\\
    c17 & 315.853497 & 68.132781 & 2270055421855682688	 & 318$^{+8}_{-8}$      & 7.334$\pm$0.146   & -1.539$\pm$0.137 & 16.103$\pm$0.002 &1.2&-&-& \checkmark & \checkmark &\checkmark &-&-2.39\\
    c18 & 316.150453 & 68.498418 & 2270280409423293952	 & 336$^{+5}_{-5}$      & 7.372$\pm$0.078   & -1.308$\pm$0.068 & 14.571$\pm$0.004 &1.4&M1&-13.01& \checkmark & -&\checkmark &-&-0.80\\
    c19 & 316.151474 & 68.496975 & 2270280405127009024	 & 366$^{+19}_{-17}$    & 7.148$\pm$0.242    & -1.375$\pm$0.237 & 17.925$\pm$0.017 &1.0&-&-& -&- &-&-&-\\
    c20 & 316.394054 & 68.209051 & 2270068512916332672	 & 336$^{+3}_{-3}$      & 8.417$\pm$0.054   & -1.372$\pm$0.053 & 14.290$\pm$0.003 &1.2&-&-& \checkmark & -&\checkmark &-&-2.67\\ 
    \hline
    \end{tabular}\\
    \renewcommand{\arraystretch}{1.0}
    Columns 2 \& 3: 2015.5 epoch Right Ascension \& Declination of sources given by \textit{Gaia} DR2.\\
    Column 4: Distance taken from the \citet{2018AJ....156...58B} catalogue.\\
    Column 9: Renormalised Unit Weight Error (RUWE) of the comoving sources.\\ 
    Column 15: X-ray detection by \textit{XMM-Newton}.\\
\end{table}
\end{landscape}
\begin{table}
    \caption{X-ray sources detected by \textit{XMM-Newton} around HD 200775.}
    \label{tab:xray_table}
    \begin{center}
    \begin{tabular}{lcccccccccc} % four columns, alignment for each 
    \hline
    $\#$ & RA (2000) & Dec (2000) & r & HR1 & e\_HR1 & HR2 & e\_HR2& 2MASS &\textit{Spitzer}&\textit{WISE}\\
     & ($^{\circ}$) & ($^{\circ}$) & ($^{\prime\prime}$) & & & & & & &\\
    \hline 
    \multicolumn{11}{c}{X-ray sources associated with the known YSO candidates} \\
    4  & 315.360137 & 68.177384 & 0.346 & 0.96 & 0.04 & 0.30 & 0.05 & \checkmark & \checkmark & \checkmark\\
    8  & 315.403845 & 68.163229 & 0.079 & 0.85 & 0.01 & 0.30 & 0.01 & \checkmark & - &-\\
    11 & 315.558856 & 68.232957 & 0.973 & 0.97 & 0.65 & 0.93 & 0.09 & \checkmark & \checkmark & \checkmark\\
    13$^{\dagger}$  & 315.617351 & 68.058043 & 0.723 & 0.79 & 0.15 & 0.24 & 0.14 & - & \checkmark &\checkmark\\
    14 & 315.637078 & 68.124783 & 0.561 & 0.72 & 0.06 &-0.10 & 0.06 & \checkmark & \checkmark & \checkmark\\
    17 & 315.849525 & 67.984878 & 2.996 & 0.92 & 0.08 & 0.31 & 0.10 & \checkmark & \checkmark &\checkmark\\
    27 & 315.133771 & 68.213195 & 0.324 & 1.00 & 0.04 & 0.73 & 0.03 & \checkmark & \checkmark & \checkmark\\
    30 & 315.261981 & 68.218805 & 0.728 & 0.94 & 0.17 & 0.47 & 0.13 & \checkmark & \checkmark & -\\
    32 & 315.286172 & 68.214544 & 0.234 & 0.83 & 0.33 & 0.67 & 0.18 & \checkmark & \checkmark & \checkmark\\
    39 & 315.406921 & 68.191929 & 0.646 & 0.94 & 0.08 & 0.21 & 0.11 & \checkmark & - & \checkmark\\
    \hline
    \multicolumn{11}{c}{X-ray sources associated with the newly identified comoving sources} \\
    c5 & 315.271092 & 68.072880 & 0.858 & 0.69 & 0.03 &-0.08 & 0.03 & \checkmark & - & \checkmark\\
    c8 & 315.421356 & 68.182647 & 1.375 & 0.81 & 0.18 & 0.19 & 0.17 & \checkmark & -& \checkmark\\
    c9 & 315.439065 & 68.113329 & 0.081 & 0.88 & 0.04 & 0.06 & 0.05 & \checkmark & \checkmark &\checkmark\\
    c14 & 315.501365 & 68.192470 & 0.123 & 0.98 & 0.09 & 0.44 & 0.10 & \checkmark & \checkmark & \checkmark\\
    \hline
    \multicolumn{11}{c}{X-ray sources not associated with any of the known YSO candidates and comoving sources} \\
    x1 & 315.169421 & 68.202133 & -     & -1.00    & 0.64     & 1.00     &0.14      &   -        &      -    & -\\
    x2 & 315.376244 & 68.176781 & 0.243 & 0.63     &0.17      & 0.43     &0.12      & \checkmark & \checkmark & \checkmark\\
    x3$^{\ddagger}$  & 315.490013& 68.036100 & 0.461 & 0.48& 0.09 & -0.54 &0.10     & \checkmark & - & -\\
    x4 & 315.722487 & 68.033205 & 0.651 & 0.44      &0.10     & -0.39    &0.09      & \checkmark & \checkmark & \checkmark\\
    x5 & 315.736835 & 68.128639 & -     & 1.00      &0.27     & 0.80     &0.08      & -          & -& -\\
    \hline
    \end{tabular}
    \end{center}
	r is the separation of the position of sources detected in \textit{XMM-Newton} and 2MASS.\\
    $^{\dagger}$ Star \#13 is a double star (the second component is identified as 13b, see Table \ref{tab:YSO_gaia}) of $\sim2^{\prime\prime}$ separation as resolved by the \textit{Gaia} DR2. However, in both \textit{Spitzer} \& \textit{WISE}, they are detected as a single source. The \textit{Spitzer} and the \textit{WISE} values are assigned to the brightest of them. \\
    $^{\ddagger}$ There are two sources within a search radius of $5^{\prime\prime}$ from the X-ray detection. 
\end{table}
\begin{landscape}
    \begin{table}  
	{\scriptsize
	\caption{2MASS, {\it Spitzer} and {\it WISE} magnitudes for the known YSO candidates and newly identified comoving sources with photometric quality `A' in all bands.}\label{tab:2mass_wise_yso_com}  
	\begin{tabular}{llcccccccccccc} % four columns, alignment for each
		\hline
		$\#$ & RA & Dec & J$\pm$eJ & H$\pm$eH & K$\pm$eK & I1$\pm$eI1 & I2$\pm$eI2 & I3$\pm$eI3 & I4$\pm$eI4 & W1$\pm$eW1 & W2$\pm$eW2 & W3$\pm$eW3 & Class\\\hline
        \multicolumn{13}{c}{Previously known YSO candidates}\\
        1 & 314.845315 & 68.245467 & 10.588$\pm$0.026 & 9.342$\pm$0.032 & 8.532$\pm$0.021 & 7.267$\pm$0.056 & 6.722$\pm$0.056 & 6.246$\pm$0.054 & 5.270$\pm$0.052 & 7.601$\pm$0.031 & 6.877$\pm$0.020 & 4.690$\pm$0.014 &II\\
        2 & 315.084447 & 68.140772 & 12.472$\pm$0.024 & 11.165$\pm$0.027 & 10.518$\pm$0.021 & 9.579$\pm$0.053 & 9.028$\pm$0.052 & 8.538$\pm$0.052 & 7.835$\pm$0.051 & 9.886$\pm$0.023 & 9.137$\pm$0.020 & 6.819$\pm$0.020 &II\\  
        3 & 315.231481 & 68.190885 & 12.831$\pm$0.024 & 12.166$\pm$0.029 & 11.749$\pm$0.020 & 11.243$\pm$0.052 & 10.906$\pm$0.052 & 10.544$\pm$0.053 & 9.711$\pm$0.052 & 11.418$\pm$0.023 & 10.879$\pm$0.021 & - &II\\  
		4$^{\dagger}$ & 315.359984 & 68.177338 & 11.107$\pm$0.042 & 10.084$\pm$0.046 & 9.571$\pm$0.039 & 9.192$\pm$0.059 & 8.250$\pm$0.053 & 7.874$\pm$0.053 & 6.607$\pm$0.075 & 8.360$\pm$0.020 & 7.647$\pm$0.018	& 4.880$\pm$0.015&II\\
		5$^{\dagger}$ & 315.362884 & 68.177214 & 12.323$\pm$0.031 & 11.150$\pm$0.036 & 10.417$\pm$0.025 & - & - & - & - & - & - & - & II\\
		6$^{\ddagger}$ & 315.399160 & 68.223752 & 12.471$\pm$0.029 & 11.033$\pm$0.038 & 10.257$\pm$0.023 & 9.139$\pm$0.054 & 8.579$\pm$0.056 & 8.536$\pm$0.052 & 7.538$\pm$0.053 & 9.355$\pm$0.023 & 8.649$\pm$0.020 & - &II\\
		7 & 315.400352 & 68.139576 & - & - & -& 10.166$\pm$0.054 & 9.404$\pm$0.053 & 8.988$\pm$0.060 & 7.663$\pm$0.087 & 9.844$\pm$0.023 & 9.114$\pm$0.020 & - & Flat\\
		8 & 315.403923 & 68.163263 & 6.111$\pm$0.030 & 5.465$\pm$0.031 & 4.651$\pm$0.017 & - & - & - &-& - & - &-& -\\
		9 & 315.427117 & 68.215960 & 11.911$\pm$0.031 & 10.892$\pm$0.033 & 10.421$\pm$0.023 & 9.752$\pm$0.056 & 9.466$\pm$0.059 & 9.090$\pm$0.053 & 8.388$\pm$0.054 & 9.905$\pm$0.023 & 9.400$\pm$0.020 & - & II\\
        10 & 315.444875 & 68.145894 & 11.792$\pm$0.024 & 10.798$\pm$0.033 & 10.159$\pm$0.022 & 9.561$\pm$0.052 & 9.246$\pm$0.051 & 9.144$\pm$0.052 & 8.760$\pm$0.054 & - & - & - & II\\   
        11 & 315.558650 & 68.233141 & 13.982$\pm$0.032 & 12.405$\pm$0.033 & 11.686$\pm$0.024 & 10.870$\pm$0.052 & 10.513$\pm$0.058 & 10.179$\pm$0.055 & 9.496$\pm$0.053 & 11.065$\pm$0.023 & 10.513$\pm$0.020 & - & II\\
        12$^{\ddagger}$ & 315.584979 & 68.423341 & 11.513$\pm$0.027 & 10.529$\pm$0.033 & 9.880$\pm$0.024 & - & - & - & - & 9.346$\pm$0.023 & 8.765$\pm$0.020 & 6.386$\pm$0.014 &II\\
        13$^{*}$ & 315.617758 & 68.058287 & - & - & - & 9.350$\pm$0.055	& 9.001$\pm$0.053 & 8.526$\pm$0.066 & 7.495$\pm$0.054 & 9.103$\pm$0.022 & 8.630$\pm$0.020 & 6.637$\pm$0.016 & II\\
        13b$^{*}$ & 315.618244 & 68.057669 & - & - & - & - & - & - & - & - & - & - &-\\
        14 & 315.637634 & 68.124746 & 11.559$\pm$0.027 & 10.713$\pm$0.029 & 10.411$\pm$0.021 & 9.614$\pm$0.054 & 9.282$\pm$0.053 & 8.937$\pm$0.053 & 8.347$\pm$0.052 & 9.905$\pm$0.023 & 9.452$\pm$0.020 & 8.021$\pm$0.028 &II\\
        15 & 315.747855 & 68.108939 & 13.871$\pm$0.035 & 13.083$\pm$0.033 & 12.357$\pm$0.028 & 10.772$\pm$0.055 & 9.968$\pm$0.053 & 9.069$\pm$0.052 & 7.279$\pm$0.054 & 11.302$\pm$0.023 & 10.105$\pm$0.020 & 5.962$\pm$0.015 &I\\
        16$^{\ddagger}$ & 315.748555 & 68.136623 & 12.414$\pm$0.029 & 11.440$\pm$0.031 & 11.066$\pm$0.023 & 10.748$\pm$0.054 & 10.626$\pm$0.054 & 10.559$\pm$0.052 & 10.449$\pm$0.057 & 10.866$\pm$0.023 & 10.643$\pm$0.020 & 10.175$\pm$0.086&III\\
        17 & 315.851719 & 67.985134 & 9.538$\pm$0.027 & 8.767$\pm$0.032 & 8.196$\pm$0.02 & 7.393$\pm$0.053 & 6.988$\pm$0.076 & 6.534$\pm$0.055 & 5.615$\pm$0.054 & 7.632$\pm$0.030 & 6.900$\pm$0.020 & 4.662$\pm$0.014 &II\\
        18$^{\ddagger}$ & 315.923249 & 68.396018 & 13.762$\pm$0.029 & 13.020$\pm$0.037 & 12.603 $\pm$0.030 & - & - & - & - & 12.189$\pm$0.023 & 11.762$\pm$0.020 & 10.090$\pm$0.035 &II\\
        19 & 315.997585 & 67.824847 & 12.789$\pm$0.029 & 11.981$\pm$0.031 & 11.478$\pm$0.019 & 10.618$\pm$0.055 & 10.225$\pm$0.052 & 9.909$\pm$0.052 & 9.284$\pm$0.053 & 10.838$\pm$0.023 & 10.231$\pm$0.020 & 8.685$\pm$0.021 &II\\
        \hline
        \multicolumn{13}{c}{Newly identified comoving sources}\\
        %c1 & 314.358914 & 68.184721 & 12.338$\pm$0.024 & 11.695$\pm$0.029 & 11.535$\pm$0.023 & 11.437$\pm$0.023 & 11.464 $\pm$0.020 & 11.434$\pm$0.096 \\
        c1 & 314.358914 & 68.184721 & 12.338$\pm$0.024 & 11.695$\pm$0.029 & 11.535$\pm$0.023& - & -&-&-& 11.437$\pm$0.023 & 11.464$\pm$0.020 & 11.434$\pm$0.096 & III\\ 
        c2 & 314.877400 & 67.678993 & 12.772$\pm$0.025 & 12.145$\pm$0.030 & 11.769$\pm$0.026& - & -&-&-& 11.531$\pm$0.023 & 11.201$\pm$0.020 & 11.006$\pm$0.069 & III\\
        c3 & 314.963955 & 68.336928 & 12.954$\pm$0.026 & 12.173$\pm$0.030 & 11.808$\pm$0.025 & - & - & - & - & 11.674$\pm$0.023 & 11.397$\pm$0.020 & 11.316$\pm$0.082 &III\\
        c4 & 315.201051 & 68.415148 & 11.715$\pm$0.026 & 10.773$\pm$0.030 & 10.495$\pm$0.019 & - & - & - & - & 10.386$\pm$0.023 & 10.349$\pm$0.020 & 10.172$\pm$0.041 &III\\
        c5 & 315.271845 & 68.072754 & 10.630$\pm$0.024 & 9.928$\pm$0.032 & 9.683$\pm$0.022 & - & - & - & - & 9.567$\pm$0.023 & 9.472$\pm$0.020 & - &III\\
        c6 & 315.350786 & 68.150243 & 13.516$\pm$0.024 & 12.851$\pm$0.029 & 12.411 $\pm$0.022 & - & - & - & - & - & - & -&-\\
        c7 & 315.420625 & 68.301379 & 12.662$\pm$0.027 & 12.051$\pm$0.033 & 11.796$\pm$0.021 & 11.498$\pm$0.051 & 11.455$\pm$0.039 & 11.851$\pm$0.086 & 11.576$\pm$0.037 & 11.643$\pm$0.023 & 11.410$\pm$0.020 & - &III\\
        c8 & 315.422157 & 68.182326 & 12.300$\pm$0.029 & 11.373$\pm$0.033 & 10.854$\pm$0.022 & - & - & - & - & 9.575$\pm$0.020 & 9.220$\pm$0.019 & - & II\\
        c9 & 315.439309 & 68.113264 & 10.426$\pm$0.026 & 9.771$\pm$0.032 & 9.613$\pm$0.022 & 9.601$\pm$0.050 & 9.606$\pm$0.051 & 9.553$\pm$0.048 & 8.616$\pm$0.040 & 9.374$\pm$0.024 & 9.362$\pm$0.020 & - &III\\
        c10 & 315.450292 & 68.115938 & 13.893$\pm$0.027 & 13.136$\pm$0.033 & 12.739$\pm$0.025 & 12.217$\pm$0.094 & 11.903$\pm$0.042 & 10.789$\pm$0.064 & 9.414$\pm$0.050 & 11.447$\pm$0.025 & 11.002$\pm$0.022 & - &II\\
        c11 & 315.452017 & 68.135740 & 14.588$\pm$0.035 & 13.876$\pm$0.048 & 13.491 $\pm$0.042 & - & - & - & - & - & - & -&-\\
        c12 & 315.458232 & 68.158415 & 12.589$\pm$0.024 & 11.948$\pm$0.032 & 11.573$\pm$0.023 & - & - & - & - & - & - & - &-\\
        c13 & 315.470563 & 68.362746 & 13.771$\pm$0.029 & 12.561$\pm$0.035 & 12.104$\pm$0.026 & 11.866$\pm$0.056 & 11.861$\pm$0.056 & 11.751$\pm$0.081 & 12.039$\pm$0.053 & 11.941$\pm$0.022 & 11.700$\pm$0.020 & 11.168$\pm$0.075 &III\\
        c14 & 315.501565 & 68.192440 & 11.968$\pm$0.027 & 10.816$\pm$0.032 & 10.387 $\pm$0.019 & 10.257$\pm$0.056 & 10.161$\pm$0.050 & 10.123$\pm$0.053 & 10.217$\pm$0.045 & 9.939$\pm$0.022 & 9.748$\pm$0.020 & - &III\\
        c15 & 315.612647 & 68.015216 & 13.258$\pm$0.030 & 12.073$\pm$0.032 & 11.585$\pm$0.024 & - & - & - & - & 11.396$\pm$0.023 & 11.107$\pm$0.020 & - & III\\
        c16 & 315.675742 & 68.172632 & 14.368$\pm$0.035 & 13.549$\pm$0.037 & 13.125$\pm$0.024 & 12.465$\pm$0.052 & 12.217$\pm$0.052 & 12.004$\pm$0.084 & 10.621$\pm$0.050 &  12.366$\pm$0.023 & 12.051$\pm$0.023 & - &II\\
        c17 & 315.853497 & 68.132781 & 12.364$\pm$0.029 & 11.410$\pm$0.030 & 11.024$\pm$0.021 & 10.885$\pm$0.051 & 10.803$\pm$0.051 & 10.689$\pm$0.050 & 10.683$\pm$0.044 & 10.881$\pm$0.022 & 10.699$\pm$0.020 & - &III\\
        c18 & 316.150453 & 68.498418 & 11.570$\pm$0.030 & 10.782$\pm$0.037 & 10.454$\pm$0.023 & - & - & - & - & 9.746$\pm$0.023 & 9.338$\pm$0.020 & 7.241$\pm$0.016 &II\\
        c19 & 316.151474 & 68.496975 & - & - & - & - & - & - & - & - & - & - &-\\
        c20 & 316.394054 & 68.209051 & 11.600$\pm$0.027 & 10.918$\pm$0.031 & 10.626 $\pm$0.021 & - & - & - & - & 10.559$\pm$0.022 & 10.511$\pm$0.020 & 10.342$\pm$0.042 & III\\
      \hline
	\end{tabular}\\
	$^{\dagger}$ Stars \#4 and \#5 are resolved as two sources in 2MASS, \textit{Spitzer} and \textit{Gaia} DR2, but \textit{WISE} detected them as a single source. We assigned the \textit{WISE} to the brightest of them in $K_{s}$.\\
    $^{*}$ Stars \#13 \& \#13b appear as double sources of $\sim2^{\prime\prime}$ in \textit{Gaia} DR2. 2MASS classified it as an extended source. Visual inspection also show them as two sources. But detected as single source by 2MASS and \textit{WISE}.\\
    $^{\ddagger}$ Stars 6, 12, 16 and 18 are identified as outliers based on our distance estimation.}
\end{table}
\end{landscape}
\begin{landscape}
    \begin{table}  
	{\scriptsize
	\caption{2MASS, {\it Spitzer} and {\it WISE} magnitudes for the rest of known YSO candidates %and X-ray sources 
	with photometric quality `A' in all bands.}\label{tab:NRI_other}  
	\begin{tabular}{lccccccccccccc} % four columns, alignment for each
	\hline
	%{\footnotesize
	$\#$ & RA & Dec & J$\pm$eJ & H$\pm$eH & K$\pm$eK & I1$\pm$eI1 & I2$\pm$eI2 & I3$\pm$eI3 & I4$\pm$eI4 & W1$\pm$eW1 & W2$\pm$eW2 & W3$\pm$eW3 & Class\\\hline
    %\multicolumn{13}{c}{Previously known YSO candidates}\\
    20 & 314.137917 & 68.246667 & 14.583$\pm$0.077 & 13.669$\pm$0.087 & 13.275$\pm$0.072 & - & - & - & - & 12.204$\pm$0.023 & 12.118$\pm$0.021 & 9.304$\pm$0.026 & II\\
    21 & 315.078827 & 68.185074 & 15.386$\pm$0.058 & 14.533$\pm$0.065 & 13.994$\pm$0.066 & - & - & - & - & 13.749$\pm$0.026 & 13.681$\pm$0.029 & - & III\\
    22 & 315.086250 & 68.221111 & - & - & - & 12.255$\pm$0.059 & 10.343$\pm$0.058 & 9.331$\pm$0.051 & 8.501$\pm$0.051 & 12.086$\pm$0.023 & 9.698$\pm$0.020 & 7.528$\pm$0.018 & I\\
    23 & 315.089576 & 68.129234 & - & - & - &- & - & - & - &- & - &- &-\\
    24 & 315.092083 & 68.216111 & - & - & - & 11.207$\pm$0.063 & 9.391$\pm$0.055 & 8.462$\pm$0.051 & 7.576$\pm$0.051 & 11.595$\pm$0.027 & 9.086$\pm$0.021 & 7.028$\pm$0.017 & I\\
    25 & 315.093333 & 68.217778 & - & - & - & 12.755$\pm$0.069 & 11.520$\pm$0.059 & 10.883$\pm$0.053 & 10.089$\pm$0.053 & - & - & -& Flat\\
    26 & 315.120417 & 68.217222 & 15.998$\pm$0.080 & 13.096$\pm$0.030 & 11.584$\pm$0.023 & 10.270$\pm$0.055 & 9.756$\pm$0.053 & 9.321$\pm$0.051 & 8.585$\pm$0.051 & 10.587$\pm$0.023 & 9.638$\pm$0.020 & 8.006$\pm$0.022 & II\\
    27 & 315.133750 & 68.213056 & 11.377$\pm$0.024 &  9.836$\pm$0.028 &  9.096$\pm$0.021 & 8.379$\pm$0.052 & 8.053$\pm$0.050 & 7.688$\pm$0.052 & 7.143$\pm$0.053 & 8.712$\pm$0.022 & 8.143$\pm$0.020 & 6.690$\pm$0.016 & II\\
    28 & 315.152500 & 68.062222 & 11.646$\pm$0.022 & 10.826$\pm$0.027 & 10.608$\pm$0.023 & 10.391$\pm$0.056 & 10.394$\pm$0.052 & 10.318$\pm$0.052 & 10.191$\pm$0.051 & 10.458$\pm$0.023 & 10.377$\pm$0.020 & - & III\\
    29 & 315.242083 & 68.160556 & 12.389$\pm$0.026 & 11.535$\pm$0.031 & 11.081$\pm$0.020 & 10.539$\pm$0.051 & 10.225$\pm$0.052 & 9.802$\pm$0.055 & 8.935$\pm$0.051 & 10.623$\pm$0.023 & 10.116$\pm$0.021 & - & II\\
    30 & 315.262500 & 68.218611 & 14.118$\pm$0.040 & 12.755$\pm$0.043 & 11.969$\pm$0.029 & 11.028$\pm$0.050 & 10.448$\pm$0.055 & 10.165$\pm$0.052 & 9.293$\pm$0.053 & - & - &- &II\\
    31 & 315.265417 & 68.219167 & - & - & - & 10.171$\pm$0.054 & 9.761$\pm$0.054 & 9.432$\pm$0.051 & 8.733$\pm$0.053 & 9.877$\pm$0.023 & 9.305$\pm$0.020 & - & II\\
    32 & 315.286250 & 68.214444 & 14.736$\pm$0.034 & 13.627$\pm$0.037 & 12.902$\pm$0.032 & 12.189$\pm$0.052 & 11.764$\pm$0.052 & 11.317$\pm$0.054 & 10.568$\pm$0.069 & 12.385$\pm$0.024 & 11.675$\pm$0.022 & - & II\\
    33 & 315.292500 & 68.192500 & 14.764$\pm$0.036 & 14.097$\pm$0.041 & 13.677$\pm$0.047 & 13.091$\pm$0.053 & 12.779$\pm$0.055 & 12.384$\pm$0.060 & 11.660$\pm$0.071 & - & - & - & II\\
    34 & 315.302083 & 68.171944 & 11.829$\pm$0.027 & 10.836$\pm$0.035 & 10.391$\pm$0.022 & 9.786$\pm$0.054 & 9.542$\pm$0.056 & 9.098$\pm$0.054 & 8.202$\pm$0.058 & 9.839$\pm$0.022 & 9.386$\pm$0.020 & - & II\\
    35 & 315.363750 & 68.193889 & 15.378$\pm$0.059 & 13.15$\pm$0.037 & 11.806$\pm$0.025 & 10.578$\pm$0.053 & 10.093$\pm$0.053 & 9.508$\pm$0.054 & 8.667$\pm$0.050 & 10.728$\pm$0.023 & 9.842$\pm$0.020 & -& II\\
    36 & 315.365622 & 68.136520 & 13.999$\pm$0.024 & 13.195$\pm$0.031 & 12.663$\pm$0.029 & - & - & - & - & 11.455$\pm$0.036 & 10.664$\pm$0.029 & -& II\\
    37 & 315.373750 & 68.229444 & 14.116$\pm$0.039 & 12.398$\pm$0.038 & 11.410$\pm$0.022 & 10.442$\pm$0.052 & 10.023$\pm$0.056 & 9.520$\pm$0.049 & 8.856$\pm$0.053 & 10.852$\pm$0.023 & 10.086$\pm$0.021 & -& II\\
    38 & 315.386667 & 68.188889 & - & - & - & 12.404$\pm$0.057 & 10.778$\pm$0.057 & 9.726$\pm$0.052 & 8.599$\pm$0.061 & 11.507$\pm$0.022 & 9.835$\pm$0.019 & -& I\\
    %7, 38 & 315.400417 & 68.139444 &  &  &  & 9.844$\pm$0.023 & 9.114$\pm$0.020 & 5.656$\pm$0.053\\
    39 & 315.406537 & 68.191931 & 11.271$\pm$0.024 & 10.260$\pm$0.031 &  9.683$\pm$0.022 & - & - & - & - & 8.935$\pm$0.022 & 8.474$\pm$0.020 & - & II\\
    40 & 315.431667 & 68.160000 & 11.218$\pm$0.024 & 10.415$\pm$0.030 & 10.156$\pm$0.020 & 9.988$\pm$0.054 & 9.951$\pm$0.052 & 9.859$\pm$0.058 & 9.961$\pm$0.075 & - & - & - &III\\
    41 & 315.432083 & 67.840556 & 13.056$\pm$0.027 & 11.82$\pm$0.031 & 11.090$\pm$0.022 & 10.300$\pm$0.051 & 9.992$\pm$0.054 & 9.756$\pm$0.053 & 9.129$\pm$0.053 & 10.478$\pm$0.023 & 9.915$\pm$0.020 & 8.335$\pm$0.021& II\\
    42 & 315.432917 & 68.234167 & 15.036$\pm$0.071 & 13.734$\pm$0.052 & 12.402$\pm$0.032 & 9.872$\pm$0.052 & 9.045$\pm$0.055 & 8.376$\pm$0.051 & 7.840$\pm$0.051 & 10.040$\pm$0.023 & 8.894$\pm$0.018 & -& I\\
    43 & 315.439167 & 68.072500 & 13.281$\pm$0.026 & 11.923$\pm$0.031 & 11.352$\pm$0.025 & 10.796$\pm$0.057 & 10.368$\pm$0.060 & 10.209$\pm$0.056 & 8.886$\pm$0.055 & 11.002$\pm$0.023 & 10.515$\pm$0.020 & -& II\\
    44 & 315.469583 & 68.164167 & - & - & - & 11.058$\pm$0.051 & 10.319$\pm$0.057 & 9.741$\pm$0.060 & 8.744$\pm$0.064 & -& -& -& II\\
    45 & 315.478158 & 68.116402 & 14.116$\pm$0.032 & 13.264$\pm$0.035 & 12.664$\pm$0.023 & - & - & - &- & 11.534$\pm$0.025 & 11.490$\pm$0.021 & -& II\\
    46 & 315.510417 & 68.199444 & 14.854$\pm$0.043 & 13.549$\pm$0.036 & 12.878$\pm$0.030 & - & - & - &- & 12.319$\pm$0.024 & 11.796$\pm$0.023 & -& Flat\\
    47 & 315.553036 & 68.397097 & 11.573$\pm$0.024 & 10.469$\pm$0.032 & 10.099$\pm$0.021 & - & - & - &- & 9.937$\pm$0.023 & 9.956$\pm$0.020 & 9.791$\pm$0.034& III\\
    48 & 315.558750 & 68.125000 & 15.768$\pm$0.075 & 14.656$\pm$0.056 & 13.604$\pm$0.044 & 11.397$\pm$0.056 & 10.438$\pm$0.054 & 9.683$\pm$0.057 & 8.435$\pm$0.064 & 11.263$\pm$0.024 & 10.167$\pm$0.021 & - & I\\
    49 & 315.588333 & 67.905556 & - & - & - & 10.772$\pm$0.055 & 9.363$\pm$0.054 & 8.566$\pm$0.053 & 7.728$\pm$0.052 & 11.389$\pm$0.022 & 9.111$\pm$0.020 & 6.931$\pm$0.016& I\\
    50 & 315.592848 & 68.203369 & - & - & - & - & - & - &- &- & - & -&-\\
    51 & 315.613750 & 67.905000 & - & - & - & 13.589$\pm$0.063 & 12.985$\pm$0.066 & 13.105$\pm$0.081 & 13.139$\pm$0.101 & - & - & -&Flat\\
    52 & 315.616667 & 68.146389 & 15.574$\pm$0.059 & 13.815$\pm$0.046 & 12.786$\pm$0.036 & 11.894$\pm$0.053 & 11.418$\pm$0.054 & 10.940$\pm$0.054 & 9.914$\pm$0.052 & - & - &-&II\\
    %13, 14 and 51 & 315.617500 & 68.058056 &  &  &  & 9.103$\pm$0.022 & 8.630$\pm$0.020 & 6.637$\pm$0.016\\
    53 & 315.624583 & 67.902222 & 15.022$\pm$0.074 & 12.035$\pm$0.038 & 10.415$\pm$0.024 & 8.621$\pm$0.054 & 7.938$\pm$0.055 & 7.392$\pm$0.051 & 6.733$\pm$0.059 & 9.131$\pm$0.023 & 7.971$\pm$0.020 & 6.271$\pm$0.015&II\\
    54 & 315.645524 & 68.200684 & 15.337$\pm$0.051 & 14.377$\pm$0.056 & 13.685$\pm$0.050 &- & - &- & - & - & - &-&-\\
    55 & 315.728750 & 68.105833 & 15.661$\pm$0.067 & 13.712$\pm$0.041 & 12.261$\pm$0.023 & 10.688$\pm$0.051 & 10.240$\pm$0.053 & 9.945$\pm$0.054 & 9.329$\pm$0.055 & 10.809$\pm$0.023 & 10.033$\pm$0.020 & - & II\\
    56 & 315.781667 & 68.142778 & 12.435$\pm$0.029 & 11.398$\pm$0.032 & 10.732$\pm$0.021 & 10.035$\pm$0.052 & 9.690$\pm$0.050 & 9.372$\pm$0.053 & 8.632$\pm$0.053 & 10.134$\pm$0.023 & 9.627$\pm$0.020 & 7.494$\pm$0.026&II\\
    57 & 315.800417 & 68.216944 & - & - & - & - & - & - & - &- & - & -&-\\
    58 & 316.065000 & 67.712778 & 6.256$\pm$0.019 &  5.068$\pm$0.018 & 4.616$\pm$0.017 & 6.100$\pm$0.178 & 4.499$\pm$0.060 & 4.042$\pm$0.071 & 3.892$\pm$0.061 & -& -& -&III\\
    \hline
    %\multicolumn{13}{c}{X-ray sources not associated with any known sources}\\
    %x1 & 315.169421 & 68.202133 & -& -& -& -& -& -&-\\
    %x2 & 315.376244 & 68.176781 & 11.147$\pm$0.027 & 10.115$\pm$0.031 & 9.602$\pm$0.023 & 8.833$\pm$0.021 & 7.999$\pm$0.017 & 3.359$\pm$0.017&II\\
    %x3$^{\dagger}$ & 315.490013 & 68.036100 & 12.841$\pm$0.033 & 12.222$\pm$0.038 & 11.924$\pm$0.029 & - & - & -&-\\
    %x4 & 315.722487 & 68.033205 & 12.083$\pm$0.027 & 11.375$\pm$0.030 & 11.230$\pm$0.021 & 11.097$\pm$0.023 & 11.086$\pm$0.020 & 10.989$\pm$0.068&TD\\
    %x5 & 315.736835 & 68.128639 & -& -& -& -& -& -&-\\
	%}
    %\hline
	\end{tabular}\\
	%$^{\dagger}$ Gaia resolved it as a double star of $\sim$1.3$^{\prime\prime}$separation.
	The classifications of the YSO candidates were taken from \cite{2009ApJS..185..198K}.\\
	$\alpha$ for \#21, \#36, \#39, \#45 and \#47 were calculated by us.\\ 
}
\end{table}
\end{landscape}

%###################################################################################################
\bsp	
\label{lastpage}
\end{document}